\documentclass[aps,prx,amsmath,twocolumn,superscriptaddress,letterpaper,floatfix]{revtex4}
\usepackage{graphicx,color}
\usepackage{verbatim}
\usepackage{amssymb}   % for math
\usepackage{amsmath}
\usepackage{amsfonts}
\usepackage{mathdots}
\usepackage{hyperref}
\usepackage{epsfig}
\usepackage{multirow}
\usepackage{braket}
\usepackage{bm}
\begin{document}
	%captionsetup[figure]{labelfont={bf},labelformat={default},labelsep=period,name={Fig.}}
	%\title{Moir\'e Ising superconductivity: theory and application to twisted transition metal dichalcogenides}
	
	\title{ Orbital Fulde-Ferrell pairing state in moir\'e Ising superconductors }
	\author{Ying-Ming Xie}	\thanks{ymxie@ust.hk}
	\affiliation{Department of Physics, Hong Kong University of Science and Technology, Clear Water Bay, Hong Kong, China}
	
	\author{K. T. Law} 	\thanks{phlaw@ust.hk}
	\affiliation{Department of Physics, Hong Kong University of Science and Technology, Clear Water Bay, Hong Kong, China}
	
	\newcommand{\YM}[1]{\textcolor{red}{[ #1]}} 
	
	\date{\today}
	
	\begin{abstract}
		In this work, we study superconducting moir\'e  homobilayer transition metal dichalcogenides where the Ising spin-orbit coupling (SOC) is much larger than the moir\'e  bandwidth. We call such noncentrosymmetric superconductors, moir\'e Ising superconductors. Due to the large Ising SOC, the depairing effect caused by the Zeeman field is negligible and the in-plane upper critical field ($B_{c2}$) is determined by the orbital effects. This allows us to study the effect of large orbital fields. Interestingly, when the applied in-plane field is larger than the conventional orbital $B_{c2}$, a finite-momentum pairing phase would appear which we call the orbital Fulde-Ferrell (FF) state. In this state, the Cooper pairs acquire a net momentum of $2q_B$ where $2q_B=eBd$ is the momentum shift caused by the magnetic field $B$ and $d$ denotes the layer separation. This orbital field-driven FF state is different from the conventional FF state driven by Zeeman effects in Rashba superconductors. Remarkably, we predict that the FF pairing would  result in a giant superconducting diode effect under electric gating when layer asymmetry is induced. An upturn of the $B_{c2}$ as the temperature is lowered, coupled with the giant superconducting diode effect,  would allow the detection of the orbital FF state.
		
		%It is important to note that the Ising SOC strongly suppresses the paramagnetic depairing effects such that this novel orbital field driven FF state can be realized experimentally.  

		%We show that in a bilayer Ising superconductor under in-plane magnetic field, strong spin-orbit coupling can suppress Zeeman effect, and the remaining orbital effect can boost intralayer Cooper pairs to finite and opposite momenta on two layers.
	\end{abstract}
	\pacs{}
	\maketitle
	
	{\emph {Introduction.}}---Since the discovery of  correlated insulating states and unconventional superconductivity in twisted bilayer graphene \cite{Cao2018, Cao2018sc},   moir\'e superlattices have become important platforms for studying  correlated physics, superconductivity, and topological states \cite{Andrei2021}. Recently, these studies have been extended to a new type of moir\'e materials based on transition metal dichalcogenides (TMD) \cite{fengcheng2018,fengcheng2019,Wang2020,Zhang2020,Fai_hubbard2020,Fengwang2020,Fai_fractional2020,Huang_fractional2021,Yangzhang2020,Fai_strip2021,Tingxin2021,Mak2022_review,Liheng2020,Das2021,Schrade2021,Dante2022,Yaohong2022,Mills2022,Yahui2022,David2022}.  Notably, the WSe$_2$ moir\'e superlattice further shows a possible signature of superconductivity, in which the resistance drops to zero at a critical temperature of about  1 to 3K \cite{Wang2020,Liheng2020}. Importantly, superconductivity appears when the Fermi energy is near the valence band top of WSe$_2$ such that the Ising spin-orbit coupling  (SOC) is exceedingly large (in the order of hundreds of meV \cite{GuiBin2013}). The Ising SOC, which pins electron spins at opposite momentum to opposite (out-of-plane) directions \cite{XiaoDi2012,Lu2015,Xi2016}, strongly suppresses the effect of in-plane Zeeman field and enhances in-plane upper critical field $B_{c2}$ \cite{Lu2015, Xi2016, Saito2016, delaBarrera2018, Ye2018, Xing2017, Sohn2018, Tongzhou2016, He2018, Yingming2020, Tewari, Manuel2017,  Aji2016, Yanase2017,Hsu2017,Mazin2020}.  Due to the large Ising SOC, the Zeeman depairing effect of the magnetic field can be ignored and the superconductivity of the moir\'e bilayer can only be suppressed by the orbital effects.

	%It is worth noting that the TMD moir\'e superlattices studied in most recent experiments are also composed of 2H-TMD layers, where the moir\'e bands originate from the top valence bands of monolayer TMDs which possess giant Ising SOC (up to hundreds of meV \cite{GuiBin2013}). 

	%In this case, the Ising SOC  is even dominant over  moir\'e bandwidth (tens of meV),  which is extremely unusual.   Such large Ising SOC   would result in strong Ising superconductivity in superconducting twisted bilayer TMD, the properties of which calls for a theoretical study.%, which we call moir\'e Ising superconductivity.
	
	%It would be interesting and desirable to know how this giant Ising SOC affects  correlated states and superconductivity in TMD moir\'e superlattices. So far,  it has been realized that this Ising SOC is crucial for determining the nature of the correlated states appearing in TMD moir\'e superlattices. For example, it has been proposed that the insulating states observed in twisted bilayer WSe2  could be a excitonic spin superfluid owing to this giant Ising SOC  []. However, the effects of this giant Ising SOC on the superconductivity for TMD moir\'e superlattices have not been discussed yet.
	%Due to the  giant Ising SOC, the cooper pairing is generally an equal mixing of spin-singlet and spin-triplet pairing.
	
	In this work, we study the role of Ising SOC in superconducting TMDs with moir\'e bands. Specifically, we show that in-plane $B_{c2}$ of the superconducting states goes beyond the Pauli limit \cite{Clogston1962,Chandrasekhar} but the in-plane $B_{c2}$ is limited by the orbital effect instead of the Zeeman effect. Moreover, we show that the moir\'e Ising superconductor can be driven to a finite-momentum pairing state at low temperatures by the orbital effects of the magnetic field. Using  realistic parameters of twisted bilayer TMDs, we find that the nature of this finite-momentum tends to be a $2q_B$-Fulde–Ferrell (FF) pairing state \cite{FF1964}, in which Cooper pairs at both layers carry a finite-momentum around $2q_B$ perpendicular to applied fields. The phase transition from the conventional pairing to the finite-momentum pairing can be detected by the temperature dependence of the upper critical field. Interestingly, we predict a giant superconducting diode effect induced by the $2q_B$-FF pairing under electric gating. The combination of $B_{c2}$ and the diode effect would provide strong evidence of the novel orbital FF state.
	%Our theory highlights moir\'e Ising superconductors, such as superconducting twisted bilayer TMDs, are a wonderful platform for exploring unconventional superconducting effects, including orbital magentic field driven  finite-momentum pairing state and gate-tunale superconducting diode effects. 

	%Finally, we  show the  nature of  finite-momentum pairing  state is highly gate tunable. Our work thus establishes the as a rather unique  Ising superconductor and the first experimentally accessible platform to explore the orbital magentic field driven  finite-momentum pairing state.

	{\emph { Model.}}---To study the properties of moir\'e Ising superconductors, we start with a continuum model of twisted homobilayer TMD with Ising SOC and  external magnetic fields \cite{fengcheng2019}.  We focus on  homobilayer TMDs  with AA stacking. The lattice constant of each monolayer is denoted by $a_0$. The top layer and the bottom layer are rotated by an angle  of $\theta/2$ and $-\theta/2$ respectively with respect to one of the transition metal sites (see Fig.~\ref{fig:fig1}(a)).   %The origin is chosen  to be on this rotational axis and halfway between layers. With respect to this point, the invariant
	The crystal point group symmetry is $D_{3}$, which is generated by a two-fold rotation $C_{2y}$ along the $y$-axis and a three-fold rotation $C_{3z}$ along the $z$-axis. It is important to note that inversion symmetry is broken in the moir\'e bilayer TMD such that the superconducting state can be different from the centrosymmetric bilayer TMD studied in Ref.\cite{Chaoxing2017}.
	% The system breaks SU(2) symmetry due to the strong Ising SOC and preserves time reversal symmetry $\hat{\mathcal{T}}$ if  the external magnetic field is absent.
	
	The  moir\`{e} superlattice, which has a moir\`{e} lattice constant of $L_M=a_0/\sin\theta$,  folds the energy bands and  gives rise to the moir\`{e} Brillouin.  The moir\`{e} bands under a finite in-plane magnetic field are described by the Hamiltonian 
	\begin{equation}\label{continuum}
		H_{\xi}(\bm{r})=\begin{pmatrix}
			h_{b}(\bm{r})&\hat{T}(\bm{r})\\
			\hat{T}^{\dagger}(\bm{r})&h_t(\bm{r})
		\end{pmatrix}.
	\end{equation}
	where $\xi=\pm$ is the valley index for $\pm\bm{K}$ valley. Here the Hamiltonian of each individual layer is given by
	\begin{eqnarray}
		h_{l}(\bm{r})&=&-\frac{1}{2m^*}(\hat{\bm{p}}+\bm{q_B}\tau_z-\xi\bm{K}_{l})^2-\mu+\Omega^{(l)}_{\xi}(\bm{r})\nonumber\\
		&&-\xi\beta_{so} s_z+u_B\bm{B\cdot s},\label{Ham2}
	\end{eqnarray}
	where $l=t(b)$ labels the top (bottom) layer,  $m^*$ denotes the  effective mass of valence band, $\mu$ is the chemical potential,  and $\tau_i$ and $s_i$ are Pauli
	matrices defined in layer and spin space, respectively. The $\beta_{so}$ characterizes the strength of Ising SOC. The orbital effect of an external  magnetic field introduces a momentum shift  $q_B=|\bm{q_B}|=eBd/2$ with  $\bm{q_B}=e\bm{A}$ and $\bm{A}=\frac{1}{2}d\bm{B}\times\bm{\hat{z}}$ as  the chosen gauge potential, where  $\bm{B}$ denotes the  in-plane external magnetic field, $d$ denotes the interlayer distance,   $e$ is the electron charge. The Zeeman effect of   the external magnetic field is captured by the last term, where the $g$ factor is taken to be 2 and  $u_B$ denotes the Bohr magneton. $\Omega_{\xi }^{(l)}(\bm{r})$ is  the intralayer moir\'e potential, and  $\hat{T}(\bm{r})$ is the interlayer moir\'e potential. The detailed form of moir\'e potentials and the model parameters adopted from Ref.~\cite{Fengwang2020} are presented in Supplementary Material (SM) Sec. I. 
	
	\begin{figure}
		\centering
		\includegraphics[width=1\linewidth]{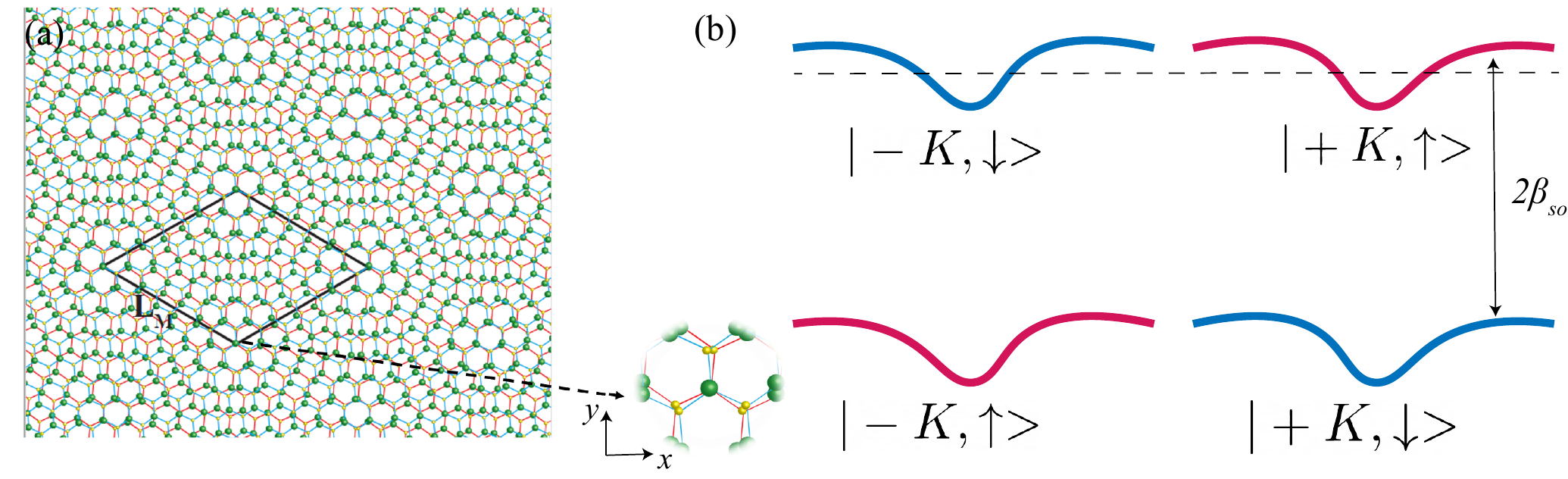}
		\caption{(a) The lattice structure of a twisted homobilayer TMD. The moir\'e unit cell is highlighted with $L_M$ as the moir\'e lattice constant. (b) A schematic plot of the top moir\'e band of spin-up state and spin-down state at two valleys. Here, $2\beta_{so}$ labels the spin-splitting induced by the Ising SOC. }%(a) Top view of the lattice structure of a twisted homobilayer 2-H TMD (tTMD) of AA stacking. The green and yellow sphere indicate the transition metal atom and chalcogen atom respectively. (b) Brillouin zones associated with the top (red) and bottom (blue) layers in a twisted bilayer, and the the moir\'e Brillouin zone (black). (c) Moir\'e bands along high symmetry lines and density of states  for tTMD at the twisted angle $\theta=4^{\circ}$. (d) Contour plot of the top moir\'{e} band in (c), the red contour line highlights the positions of half-filling.}
		\label{fig:fig1}
	\end{figure}

	We describe the superconducting twisted homobilayer TMD  by a mean-field Hamiltonian, which is written as
	\begin{equation}
		H_{MF}(\bm{r})=H(\bm{r})+\sum_{\xi} \Psi^{\dagger}_{\xi }(\bm{r}) \hat{\Delta}(\bm{r}) \Psi^{\dagger}_{-\xi}(\bm{r})+\text{H.c.}.\label{model}
	\end{equation}
	Here the moir\'{e} Hamiltonian
	\begin{equation}
		H(\bm{r})=\sum_{\xi}\int d\bm{r}\Psi^{\dagger}_{\xi}(\bm{r})\mathcal{H}_{\xi}(\bm{r})\Psi_{\xi}(\bm{r})\label{continue_Ha}
	\end{equation} 
	and  $\Psi_{\xi}(\bm{r})=(\psi_{\xi b\uparrow},\psi_{\xi b\downarrow},\psi_{\xi t\uparrow},\psi_{\xi t\downarrow})^{T}$ denotes a four-component electron annihilation operator. The  pairing matrix $\hat{\Delta}(\bm{r})$ is represented in the layer and spin space.  Due to the layered structure, we expect the pairings to be within electrons of the same layer which can be  classified with irreducible representations of $D_3$ point group (see SM Sec.~I \cite{Supp}). The favoured pairing form is determined by the microscopic interaction. In this work, for illustrative purposes, we consider the two conventional gapped intralayer pairings: $\hat{\Delta}_{A_1}=\Delta is_y$ and $\hat{\Delta}_{A_2}=\Delta i\tau_zs_y$, where $A_1$, $A_2$ label the irreducible representations of $D_3$. Here, we consider both $A_1$ and $A_2$ pairings as they would generally be  mixed by in-plane magnetic fields in the case of finite-momentum pairings.

	{\emph {The enhanced in-plane upper critical field $B_{c2}$.}}---
	The in-plane $B_{c2}$ of the moi\'re Ising superconductor can be obtained from the linearized gap equation 
	\begin{equation}
		U_0\chi_s(\bm{q},\bm{B},T)=1.\label{linear}
	\end{equation}
	Here, $U_0$ denotes the interaction strength that stabilizes $A_{1(2)}$-pairing, $\bm{q}$ is to take account of the possible finite pairing momentum,  $T$ is the temperature and the superconducting susceptibility $\chi_s(\bm{q},\bm{B},T)$, in general, is given by the maximal eigenvalue of  the pairing susceptibility matrix
	\begin{equation}
		\hat{\chi}(\bm{q},\bm{B},T)=\begin{pmatrix}
			\chi_{11}(\bm{q},\bm{B},T)&\chi_{12}(\bm{q},\bm{B},T)\\\chi_{21}(\bm{q},\bm{B},T)&\chi_{22}(\bm{q},\bm{B},T)
		\end{pmatrix}.\label{pairing_ma}
	\end{equation}
	The susceptibility matrix is expressed in the  $\hat{\Delta}(\bm{q})=(\Delta_{A_1}(\bm{q}),\Delta_{A_2}(\bm{q}))^{T}$ space. More details of the calculations for the pairing susceptibility and the $B_{c2}$ from  the Hamiltonian $H_0(\bm{p},\bm{B})$ can be found in SM Sec.~VI.   To be specific, we would fix the filling at $\nu\approx-0.6$ in our calculations and set the field direction along $x$-direction. In general, a three-fold anisotropy would be expected for the upper critical field.  In the main text, we set the twist angle $\theta=5^\circ$, near where the possible signature of  superconductivity would  appear in the experiment \cite{Wang2020}.
	
	\begin{figure}
		\centering
		\includegraphics[width=1\linewidth]{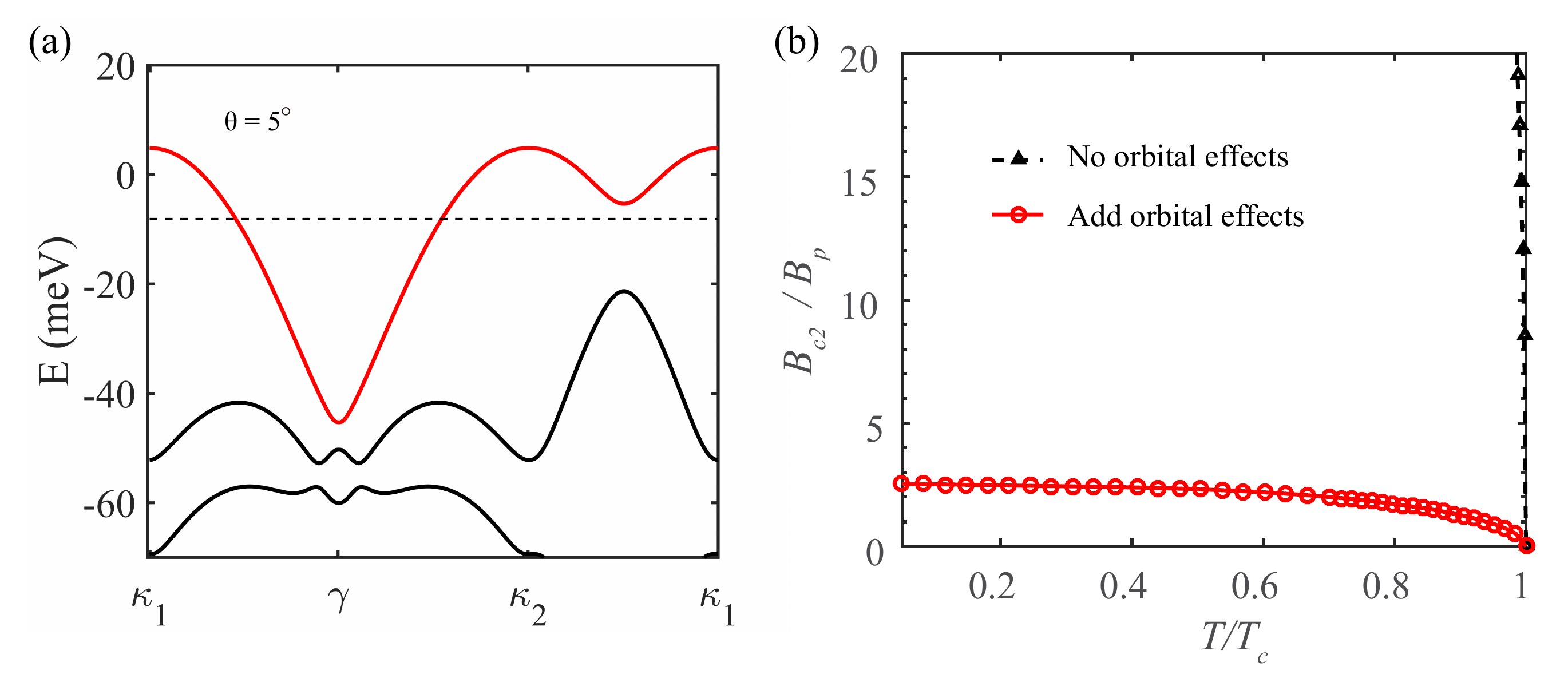}
		\caption{(a) The moir\'e bands of a homobilayer TMD with twist angle $\theta=5^{\circ}$, where the top moir\'e bands are highlighted in red. (b) The in-plane upper critical field $B_{c2}$  (in units of Pauli limit $B_p\approx1.86T_c$ ) as a function of temperature (in units of the zero-field critical temperature) with (in red) and without (in blue) orbital effects of the magnetic field. We set $T_c=1$ K and fix the chemical potential at $\nu\approx -0.6$ (the black dashed line in (a)) in (b).  }
		\label{fig:fig2}
	\end{figure}
	
	The calculated in-plane $B_{c2}$ of the zero-momentum pairing with $\bm{q}=0$ is shown in Fig.~\ref{fig:fig2}.  Figure \ref{fig:fig2}(a) displays the corresponding  moir\'e energy bands at $K$ valley, where the spin of  the top moir\'e band that contributes to the superconductivity (in red) is fully polarized by the Ising SOC. In this case, the in-plane critical magnetic field  $B_{c2}$ (in the unit of the Pauli limit $B_p$) versus critical temperature $T$ (in units of zero-field critical temperature $T_c$)  curves  are plotted in Fig.~\ref{fig:fig2}(b), where the orbital effects are present or absent according to Eq.~(\ref{Ham2}). When the orbital effects are artificially turned off while the Zeeman effects are included, it can be seen that the superconducting critical temperature is almost insensitive to the external fields due to the strong Ising SOC.  In contrast, the in-plane $B_{c2}$ would ultimately be limited to several $B_p$ when orbital effects are included (red line).  This stands in sharp contrast to superconducting MoS$_2$ and NbSe$_2$ where the depairing due to the paramagnetic effect is dominant because of the much smaller Ising SOC at the Fermi energy in these materials.

	To  estimate the magnitude of the resulting  orbital effect limited $B_{c2}$, we can construct a phenomenological GL free energy theory by taking the order parameter of top and bottom layer to be $\Delta_{t}\equiv|\Delta| e^{i\varphi_t }$ and $\Delta_{b}\equiv |\Delta| e^{i\varphi_b}$, respectively. The  Ginzburg- Landau (GL) free energy that captures our system  can be written as (see  SM Sec.V for more details):
	\begin{eqnarray}
		\mathcal{F} (|\Delta|)=\nonumber& -(\alpha_0-\alpha_1(B))|\Delta|^2+\frac{\beta_0}{2}|\Delta|^4+\\&\lambda_J(1-\cos(\varphi_t-\varphi_b))|\Delta|^2.\label{free_en}
	\end{eqnarray}
	Here, $\alpha_0\propto (T_c-T)$ and $\beta_0$ are the GL coefficients, $\alpha_1(B)$  to the second order can be approximated as $\alpha_1(B)=\Lambda q_B^2$. $\Lambda$ depends on the electron effective mass and interlayer coupling.  $\lambda_J$ denotes the Josephson coupling strength between two layers.  As expected,  the critical field $B_{c2}$ for zero-momentum pairing is determined by the $A_1$ pairing, where $\varphi_t=\varphi_b$ to  minimize the Josephson coupling energy.  According to the coefficient of $|\Delta|^2$,
	the upper critical field is now estimated as
	\begin{equation}
		B_c=\sqrt{\frac{4\alpha_0}{e^2d^2\Lambda}}.\label{upper_critical}
	\end{equation}  
	Therefore, the orbital effect limited $B_{c2} \propto \sqrt{T_c-T}$ is mainly determined by the effective mass and thickness. Note that the effective mass strongly depends on the  twist angle. As shown in SM Sec.~III, the in-plane $B_{c2}$ can be enhanced prominently when we artificially decrease the twist angle.

	{\emph {Orbital Fulde-Ferrell pairing state.}}---
	Next, we study the case of finite-momentum pairings with $\bm{q}\neq 0$ induced by the orbital effects of magnetic fields.  The stabilized finite-momentum pairing is expected to be $\bm{q}=(0,q)$, as the orbital motion of electrons is perpendicular to the in-plane magnetic fields. To find the robust $q$ driven by the in-plane magnetic fields, we display the critical field $B_c$ as a function of $q$ in Fig.~\ref{fig:fig3}(a) at various temperatures with $T=(0.9,0.5,0.1)T_c$ . Here, we have used the  magnitude of the momentum shift  $q_B=|\bm{q_B}|$ as defined in Eq.~(\ref{Ham2}) as a natural unit for the pairing momentum $q$.    The robust finite-momentum pairing can be determined by the one with $q$ that maximizes the critical field $B_{c2}$. Notably,  although the  zero-momentum pairing $q=0$ is favored near the critical temperature, a prominent $q\approx \pm2q_B$ pairing becomes favorable at low  temperatures. Figure \ref{fig:fig3}(b) displays the superconducting pairing  $\chi_s$ versus $B$ curve at $q=0$ and $|q|=2q_B$. It clearly shows that the finite-momentum pairing state with $|q|=2q_B$ exhibits a higher $B_{c2}$ than the zero-momentum pairing state.  This $2q_B$ finite-momentum pairing can be understood  from the momentum shift induced by orbital effects. The momentum of electrons at two opposite valleys, which would pair together,  obtains the same $q_B$ momentum shift according to Eq.~(\ref{Ham2}).  %Note that due to the interlayer coupling, the momentum shift near Fermi energy is not uniform (see SM Sec. III). It is particularly surprising that this 2$q_B$ finite-momentum pairing could still appear.
	
	To understand the nature of this $2q_B$-finite-momentum pairing, we can check the finite-momentum pairing susceptibility $\chi_{ij}(q=2q_B)$ versus $B$.  The stabilized pairing form could be obtained  from the pairing susceptibility matrix Eq.~(\ref{pairing_ma}), which can be written as
	\begin{eqnarray}
		\Delta(\bm{r})=\sum_{\bm{q}}\Delta_{\bm{q}}(\cos\frac{\theta_{\bm{q}}}{2}+\sin\frac{\theta_{\bm{q}}}{2}\tau_z)i\sigma_ye^{i\bm{q}\cdot\bm{r}}\label{p1} \label{p2}
	\end{eqnarray}
	Here,  $(\cos\frac{\theta_{\bm{q}}}{2},\sin\frac{\theta_{\bm{q}}}{2})^{T}$ represents the corresponding eigenvector of  $\chi_s$ with $\theta_{\bm{q}}=\arcsin  \frac{\chi_{12}}{\sqrt{(\chi_{11}-\chi_{22})^2/4+\chi_{12}^2}}$.%, and $N$ is a normalization factor to ensure the average gap of different order parameter to be the same.  
	\begin{figure}
		\centering
		\includegraphics[width=1\linewidth]{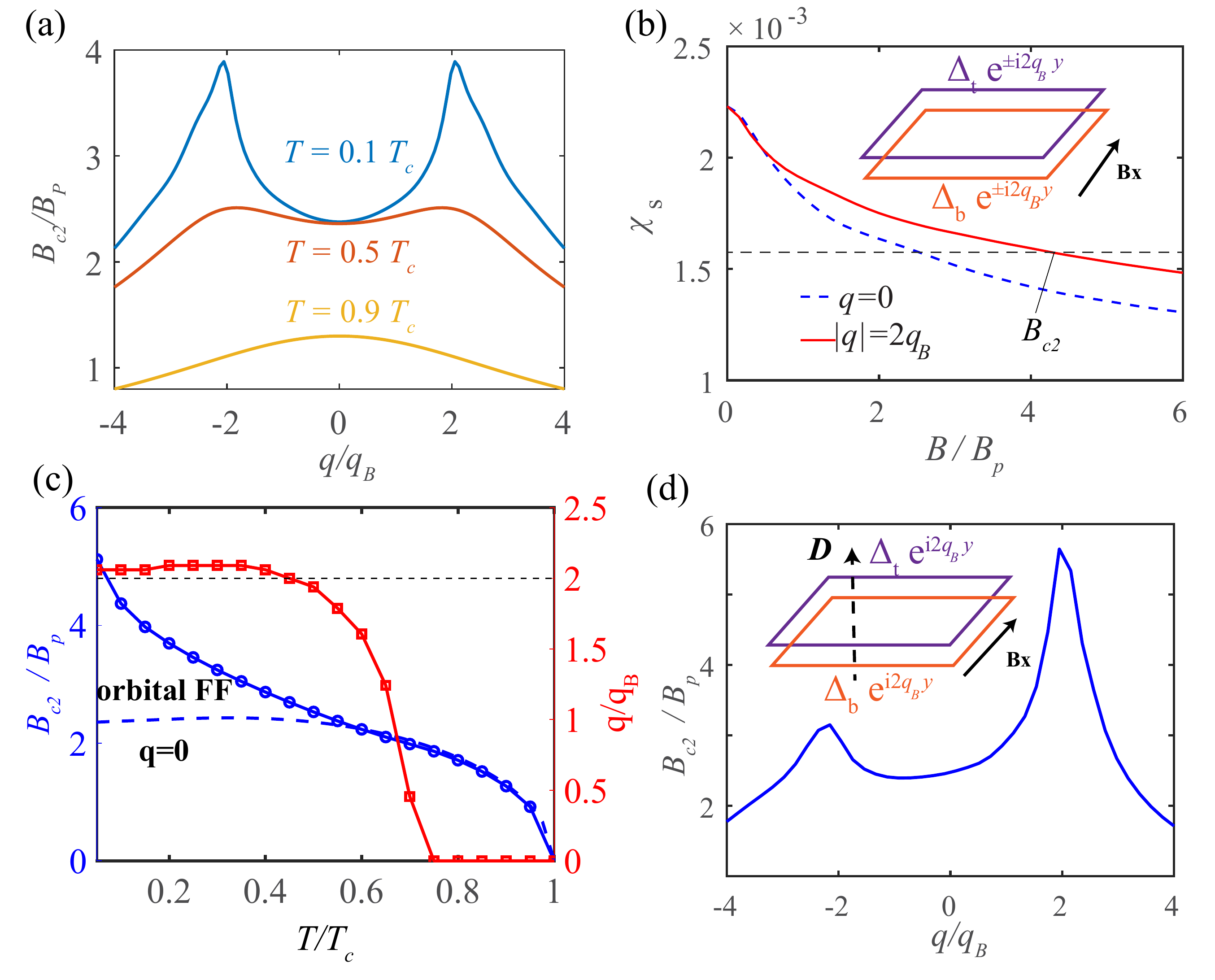}
		\caption{(a)The in-plane $B_{c2}$ versus pairing momentum $q$ (in units of $q_B$) at various temperatures $T=0.1 T_c, 0.5 T_c, 0.9 T_c$. (b) The superconducting pairing susceptibility $\chi_s (q=0)$ and $\chi_s(q=2q_B)$ versus $B$, obtained from diagonalizing  the pairing susceptibility matrix. The inset schematically plots the FF pairing with $q_y=2q_B$ driven by an in-plane field $B_x$. (c) The left axis shows the $B_{c2}$ versus $T$ curve for finite-momentum $q$ pairing (solid blue line) and zero-momentum pairing ($q=0$), while the right axis (red line) shows the corresponding  favorable pairing momentum $q$ (in units of $q_B$) as a function of temperature.  (d) The $B_{c2}$ versus $T$ curve upon a finite out-of-plane displacement field $D$.  The inset schematically represents the 2$q_B$-FF pairing under finite out-of-plane displacement fields $D$.   }
		\label{fig:fig3}
	\end{figure}
	Due to the presence of finite interlayer coupling,  the resulting finite-momentum pairing susceptibility $\chi_{11}-\chi_{22}\gg \chi_{12}$    so that $\theta_{\bm{q}}\approx 0$ (SM Sec.~III). As a result, according to Eq.~(\ref{p1}),  the stabilized  pairing form behaves as a FF pairing, which can be parameterized as $\Delta(\bm{r})=|\Delta|e^{i\bm{q}\cdot\bm{r}}$ or $\Delta(\bm{r})=|\Delta|e^{-i\bm{q}\cdot\bm{r}}$ with $q=(0,2q_B)$ (see an illustration in the inset of Fig.~\ref{fig:fig3}). We denote these two pairings as $\pm$ 2$q_B$-FF pairings. Note that although these two pairings with opposite pairing momentum are nearly degenerate, the mixing of them is not favorable according to the GL free energy analysis up to the fourth order (see SM Sec.~V). 
	Moreover, according to a phenomenological GL analysis in Sec.~V, the interlayer coupling would increase the kinetic energy of the superconductor under in-plane magnetic fields due to the canonical momentum mixing between the two layers. On the other hand, the  $2q_B$-FF pairing would lower this energy, which could make it more favorable than the zero-momentum pairing.
	%The $\pm2q_B$-FF pairings would  save the the kinetic  energy arising canonical momentum mixing of two layers comparing to the zero momentum case, but will be ultimately suppressed by increasing of kinetic energy arising from intralayer  canonical momentum.   

	To obtain the $B-T$ phase diagram,  we plot the critical $B_{c2}$ (left axis, solid blue) and the corresponding stabilized $\bm{q}=(0,q)$ (right axis, red) as a function of temperature $T$ in Fig.~\ref{fig:fig3}(c). The finite-momentum pairing $(q>0)$ is seen to emerge at temperature $T\approx 0.75 T_c$,  near where the $B_c$--$T$ curve  exhibits an upturn at the phase transition. Notably,  the momentum shifts $q$ would saturate and  the previously discussed $2q_B$-FF pairings emerge at low temperatures $T\lesssim 0.5 T_c$. The finite-momentum pairing phase region, the boundary of which is roughly given by the $B_c$--$T$ curve with $q=0$ and finite $q$,   is highlighted in Fig.~\ref{fig:fig3}(c). It can be seen that the $2q_B$-FF state can be stabilized with a temperature $T\lesssim 0.5 T_c$ and a magnetic field $B$ roughly higher than $2B_p$. 
	
	Finally, we point out that the degeneracy between  $+2q_B$-FF pairing and  $-2q_B$-FF pairing can be lifted extrinsically by  out-of-plane displacement fields $D$, which induces layer asymmetry. As shown in Fig.~\ref{fig:fig3}(d), when  an out-of-plane displacement field $D=5$ meV is applied, the $B_{c2}$ of $2q_B$ finite-momentum pairing becomes much higher than the $B_{c2}$ of the $-2q_B$ pairing, implying that $+2q_B$-FF pairing would be the favorable finite-momentum pairing under a large in-plane magnetic field. Note that in the experiment, superconductivity of twisted bilayer TMDs occurs in the presence of a displacement field. %Our calculation thus concludes that the finite-momentum pairing driven by the in-plane orbital magnetic fields  in twisted bilayer TMD could be either  $+2q_B$-FF pairing or $-2q_B$-FF pairing at a low temperature region.  

	% Hence, we find the nature of finite-momentum pairing would be   $|\Delta_{\pm}|$ can be determined b where.  the stabilized pairing tends to be    using $\Delta(\bm{r})$.    To  the component of $|\Delta_{+}|$ 
	%  indicating a LO pairing with spatially dependent pairing amplitude. his LO pairing is  

	% with $\bm{q}\approx (0, 2q_B)$ , where the mixing arises from the degeneracy at $\pm 2q_B$ and the averaging pairing gap  can be obtained as  $ |\Delta|=\sqrt{|\Delta_{+}|^2+|\Delta_{-}|^2}$.  

	\begin{figure}
		\centering
		\includegraphics[width=1\linewidth]{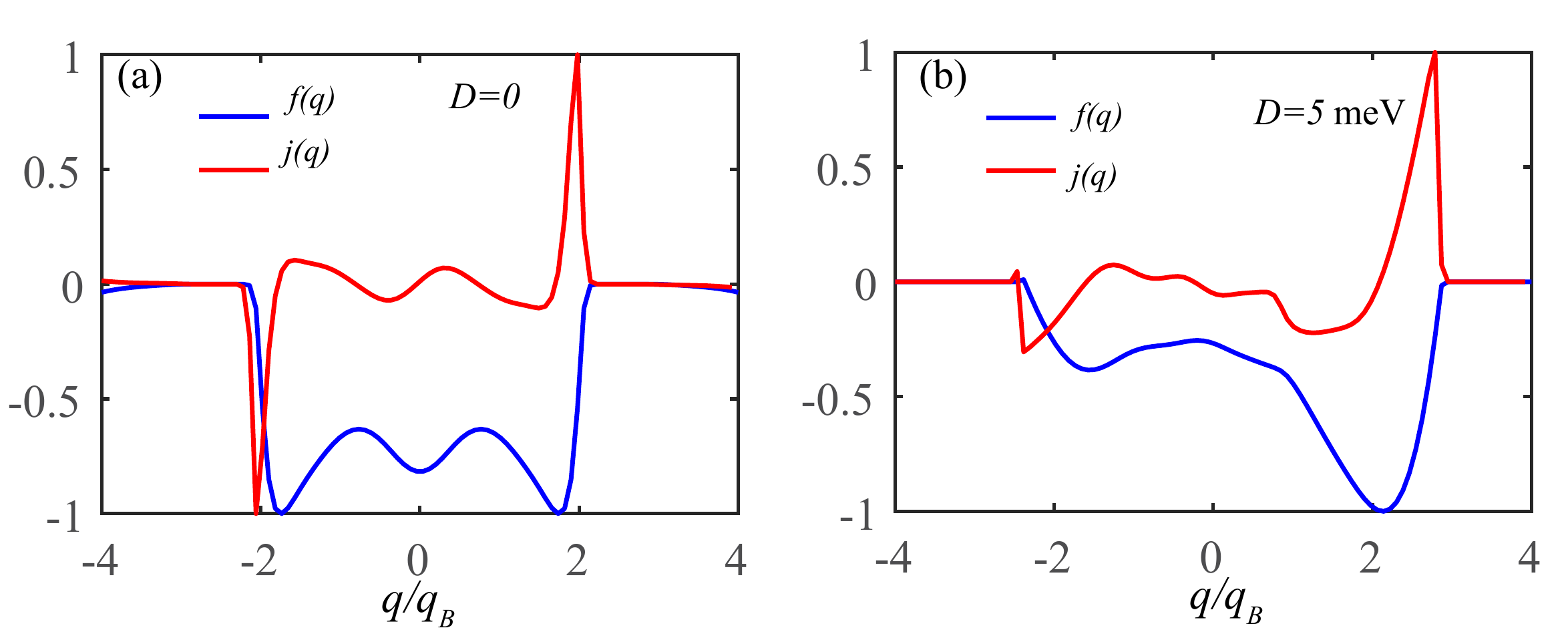}
		\caption{(a) and (b) show the free energy $f(q)$ and supercurrent $j(q)$  normalized to $[-1,1]$ without displacement fields ($D=0$ meV) and with a finite displacement field ($D=5$ meV) respectively. Here the temperature $T=0.1 T_c$, and $B=3 B_p$ and $B=2.5 B_p$ for (a) and (b), respectively. In (a), the maximum magnitudes of $j(q)$ are the same in the positive and negative directions. This indicates the absence of the superconducting diode effect. In (b),  the maximum magnitudes of $j(q)$ are different for currents flowing in opposite directions, indicating the presence of the superconducting diode effect.}
		\label{fig:fig4}
	\end{figure}
	
	{\emph {Gate-tunable superconducting diode effect.}}--- Next, we demonstrate a gate-tunable superconducting diode effect based on the proposed $2q_B$-FF pairing in moir\'e Ising superconductors. The superconducting diode effect is characterized by the critical  current difference between  currents flowing in opposite directions: $\Delta j_{c}=(j_{c,+}-|j_{c,-}|)/(j_{c,+}+|j_{c,-}|)$ \cite{Ando2020,Yanase2022,Noah2022,He_2022, Bergeret2022}, where the $+$ and $-$ signs denote the opposite current directions respectively. %This nonreciprocal response is allowed in superconductors with time-reversal and inversion symmetry broken. Hence, we also expect that the diode effect can be induced by the $2q_B$-FF pairings in moir\'e Ising superconductor. 
	
	To demonstrate this, we can calculate the supercurrent $j(\bm{q})$ from the free energy \cite{XiePRL_2020}
	\begin{equation}
		f_s(\Delta,\bm{q})=\frac{|\Delta|^2}{U_0}-\frac{1}{\beta}\sum_{\bm{k},n}\ln (1+e^{-\beta \epsilon_{n\bm{q}}(\bm{k})}).
	\end{equation}
	where $\beta=1/k_BT$, $\epsilon_{n\bm{q}}(\bm{k})$ is the quasi-particle energy of the finite-momentum Bogoliubov-de Gennes (BdG) Hamiltonian (see SM Sec.~IV \cite{Supp} for more details).
	The supercurrent $j(\bm{q})$ can be obtained by $j(\bm{q})=2\frac{\partial f(\bm{q})}{\partial \bm{q}}$, where $f(\bm{q})$ is  the lowest free energy at each pairing momentum $\bm{q}$  and is given by  minimizing  the  free energy  $f_s(\Delta, \bm{q})-f_n$ (note $f_n\equiv f_s(\Delta=0)$ is the normal state free energy) with respect to $\Delta$. Here, we consider the current direction to be along $y$-direction so that we can denote $\bm{q}=(0,q)$. 
	
	The landscape of the minimized free energy $f(q)$ (blue line) and the corresponding supercurrent $j(q)$ (red line)  in the case without displacement fields ($D=0$) and with displacement fields ($D=5$ meV) are plotted in Fig.~\ref{fig:fig4}. Here a large in-plane magnetic field ($B/B_p=3$ and $B/B_p=2.5$ for (a) and (b), respectively), and a temperature $T=0.1T_c$  are adopted so that the system is deep in the FF pairing state. It is important to note that Ising SOC is very essential here. Without Ising SOC, the superconductivity could have been killed by the paramagnetic effect before reaching the FF state. Without displacement fields (Fig.~\ref{fig:fig4}(a)), the  free energy of $q$ near $\pm 2q_B$ is lower than $q=0$ under a large $B$. In other words, $\pm 2q_B$-FF pairing would be stabilized, being  consistent with the previous linearized gap equation calculation. However, the diode effect is absent ($\Delta j_c=0$) in this case (Fig.~\ref{fig:fig4}(a)). As shown in Fig.~\ref{fig:fig4}(b), the diode effect becomes finite at finite displacement fields ($D=5$ meV). Notably, the resulting $\Delta j_c \approx 53 \% $ is much larger than the one proposed in superconductors with Rashba SOC \cite{Yanase2022}. This giant superconducting diode effect originates from the lifting of  the degeneracy between $2q_B$-FF pairing and $-2q_B$-FF pairing  by the displacement field, which enables a highly asymmetric free energy configuration as shown in Fig.~\ref{fig:fig4}(b). The implementation of an electric gate-tunable  superconducting diode effect is generally difficult in previous systems \cite{Ando2020,Wu2022, Bauriedl2022}, as the high electron density hinders the gate-controllability. The giant gate-tunable superconducting diode effect in the present system is potentially useful for dissipationless electronics, superconducting circuits and superconducting computing devices. 
	
	{\emph {Discussion.}}--- 
	It is worth noting that the pairing form $\Delta(\bm{r})$ can be changed if the interlayer coupling strength can be tuned. For example, as shown in SM Sec.~VI, we  obtained  a layer-antisymmetric FF pairing  analytically, where  $\Delta_t=|\Delta| e^{i\bm{q}\cdot\bm{r}}$ and $\Delta_b=|\Delta| e^{-i\bm{q}\cdot\bm{r}}$ with $\bm{q}=(0,2q_B)$, in the case without twisting and in the weak interlayer coupling limit.  We note that this exotic pairing has been proposed in centrosymmetric AB stacked bilayer TMDs without twisting previously \cite{Chaoxing2017}.   This layer-antisymmetric FF pairing is energetically not favored in  our case due to the  stronger interlayer coupling strength, which increases the  Josephson coupling energy. The orbital FF pairings we find  would not afford such Josephson coupling energy and are particularly allowed by noncentrosymmetric superconductors.
	
	In conclusion, we have proposed an intriguing  noncentrosymmetric superconductor---moir\'e Ising superconductor, in which the Ising SOC is dominant over moir\'e bandwidth and can be readily realized in superconducting moir\'e TMDs. We have highlighted that  moir\'e Ising superconductors are  wonderful platforms for exploring novel superconducting effects, including orbital magnetic field-driven  finite-momentum pairing state and gate-tunable superconducting diode effects. In principle, our theory for the orbital FF pairing state  can also be applied to some other non-twisted superconducting materials with inversion broken and giant Ising SOC. 
	
	%Finally, the quality of the superconductor can be critical issues.  disorder effects and
	
	{\emph { Acknowledgments.}}---K.T.L. acknowledges the support of the Ministry of Science and Technology, China, and HKRGC through Grants No. 2020YFA0309600, No. RFS2021-6S03, No. C6025-19G, No. AoE/P-701/20, No. 16310520, No. 16310219, No. 16307622 and No. 16309718. Y.M.X. acknowledges the support of HKRGC through PDFS2223-6S01. 
	
	{\emph {Note.}}--- After drafting this work, we were informed by Justin Ye that the orbital-field-driven finite-momentum pairing state might have been observed in  multilayer 2H-NbSe$_2$ \cite{YeJianting2022}.

\bibliographystyle{apsrev4-1} 
\bibliography{Reference}

	% 
	%
	%\begin{thebibliography}{99}
	%	\bibitem{Fatemi} V. Fatemi, S. Wu, Y. Cao, L. Bretheau, Q. D. Gibson, K. Watanabe, T. Taniguchi, R. J. Cava and P. Jarillo-Herrero, Science {\bf362}, 926--929 (2018).
	%	
	%\end{thebibliography}

	%
	%\begin{thebibliography}{99}
	%	\bibitem{Fatemi} V. Fatemi, S. Wu, Y. Cao, L. Bretheau, Q. D. Gibson, K. Watanabe, T. Taniguchi, R. J. Cava and P. Jarillo-Herrero, Science {\bf362}, 926--929 (2018).
	%	
	%\end{thebibliography}

	% 
	
	%
	%\begin{thebibliography}{99}
	%	\bibitem{Fatemi} V. Fatemi, S. Wu, Y. Cao, L. Bretheau, Q. D. Gibson, K. Watanabe, T. Taniguchi, R. J. Cava and P. Jarillo-Herrero, Science {\bf362}, 926--929 (2018).
	%	
	%\end{thebibliography}

		\clearpage
		\onecolumngrid
\begin{center}
		\textbf{\large Supplementary Material for\\ ``Orbital Fulde–Ferrell pairing state in Moir\'e Ising superconductors''}\\[.2cm]
		Ying-Ming Xie,$^{1}$  K. T. Law$^{1}$\\[.1cm]
		{\itshape ${}^1$Department of Physics, Hong Kong University of Science and Technology, Clear Water Water Bay,  Hong Kong, China}
\end{center}
	
	\maketitle

\setcounter{equation}{0}
\setcounter{section}{0}
\setcounter{figure}{0}
\setcounter{table}{0}
\setcounter{page}{1}
\renewcommand{\theequation}{S\arabic{equation}}
\renewcommand{\thesection}{ \Roman{section}}

\renewcommand{\thefigure}{S\arabic{figure}}
\renewcommand{\thetable}{\arabic{table}}
\renewcommand{\tablename}{Supplementary Table}

\renewcommand{\bibnumfmt}[1]{[S#1]}
\renewcommand{\citenumfont}[1]{#1}
\makeatletter

\maketitle

\setcounter{equation}{0}
\setcounter{section}{0}
\setcounter{figure}{0}
\setcounter{table}{0}
\setcounter{page}{1}
\renewcommand{\theequation}{S\arabic{equation}}
\renewcommand{\thesection}{ \Roman{section}}

\renewcommand{\thefigure}{S\arabic{figure}}
\renewcommand{\thetable}{\arabic{table}}
\renewcommand{\tablename}{Supplementary Table}

\renewcommand{\bibnumfmt}[1]{[S#1]}
\makeatletter

\maketitle

 	\section{Details for the moir\'e potential and model parameters}
 In this supplementary material section, we present the detailed form of the moir\'e potential and the adopted model parameters for the twisted homobilayer TMD in the main text. 
 The intralayer moir\'e potential is given by
 \begin{equation}
 	\Omega_{\xi s}^{(l)}(\bm{r})=V_{\xi s}\sum_{j=1,3,5}e^{i\xi(\bm{g_j\cdot r}+l\psi_{\xi s})}+h.c.,
 \end{equation}
 where $V_{\xi s}$ and $\psi_{\xi s}$ ($\xi$, $s$ are valley and spin indices), respectively, characterize the amplitude and the shape of intralayer moir\'e potential, and  the moir\'e reciprocal lattice vectors   $\bm{g_i}=\frac{4\pi}{\sqrt{3}L_M}(\cos\frac{(i-1)\pi}{3},\sin \frac{(i-1)\pi}{3})$. The interlayer tunneling  moir\'e potential $\hat{T}(\bm{r})$ in main text is given by
 \begin{eqnarray}\label{inter_couple}
 	\hat{T}(\bm{r})&=&\begin{pmatrix}
 		u_{\xi\uparrow}&u_{\uparrow\downarrow}\\
 		u_{\downarrow\uparrow}&u_{\xi\downarrow}
 	\end{pmatrix}+\begin{pmatrix}
 		u_{\xi\uparrow}&u_{\uparrow\downarrow}\omega^{-1}\\
 		u_{\downarrow\uparrow}\omega&u_{\xi\downarrow}
 	\end{pmatrix}e^{-i\xi\bm{g_2\cdot r}}\nonumber\\
 	&&+\begin{pmatrix}
 		u_{\xi\uparrow}&u_{\uparrow\downarrow}\omega\\
 		u_{\downarrow\uparrow}\omega^{-1}&u_{\xi\downarrow}
 	\end{pmatrix}e^{-i\xi\bm{g_3\cdot r}},
 \end{eqnarray}
 with $\omega=e^{i2\pi/3}$.

 To roughly capture the relevant energy scale of twisted bilayer TMD, in the calculation, we adopt the model parameters given in ref.~[5] in the main text: $1/2m^*=510.45$ meV, $\beta_{so}=110.25$ meV. $(V_{+\uparrow}, \psi_{+\downarrow}, u_{+\uparrow})=( 8 \text{ meV}, -89.8^\circ, -8.5 \text{ meV}), (V_{+\downarrow}, \psi_{+\downarrow}, u_{+\downarrow})=(7.7\text{ meV}$, $-88.35^\circ, -6.5 \text{ meV})$ and $u_{\uparrow\downarrow}=-i5.6$ \text{meV}, which are obtained by fitting the first-principle band structure of homobilayer MoTe$_2$. With the reversal symmetry operation, we have $V_{\pm\uparrow}=V_{\mp\downarrow}, \psi_{\pm\uparrow}=\psi_{\mp\downarrow}, u_{\pm\uparrow}=u_{\mp\downarrow}$.
 
 \section{ Pairing classifications for twisted bilayer TMD}
 In the absence of displacement fields and external fields, the twisted bilayer TMD respects $\hat{\mathcal{T}}\times D_3$ symmetry. Here,  $\hat{\mathcal{T}}=is_yK$ with $K$ as complex conjugate denotes time-reversal symmetry, and $D_3$  point group symmetry  is generated by a three-fold rotational symmetry  $C_{3z}=e^{-i\frac{\pi}{3}s_z}$ and an in-plane two-fold rotational symmetry $C_{2y}=-i\tau_xs_y$. 
 
 The continuum Hamiltonian for the superconducting part can be written as
 \begin{equation}
 	H_{SC}(\bm{r})=\int d\bm{r}\sum_{\xi, l,l',s,s'} \psi^{\dagger}_{\xi ls }(\bm{r}) \hat{\Delta}_{ll',ss'}(\bm{r}) \psi^{\dagger}_{-\xi l's'}(\bm{r})+\text{H.c.}.\label{SC_continuum}
 \end{equation}
 Here, $l=t,b$ is the layer indices. We can classify all possible pairings with the irreducible representations of the $D_3$ point group. This classification is done by noting (i) the pairing matrix transforms as
 \begin{eqnarray}
 	&&\hat{\mathcal{T}}: \hat{\Delta}(\bm{r})\mapsto s_y \hat{\Delta}^{*}(\bm{r}) s_y \label{trans1}\\
 	&&g: \hat{\Delta}(\bm{r})\mapsto U^{T}(g) \hat{\Delta}(g\bm{r}) U(g)\label{trans2}, 
 \end{eqnarray}
 where $g$ denotes the symmetry operation $C_{3z}$ and $C_{2y}$, $U(g)$ is the matrix representation of the symmetry operation $g$.  (ii) Due to the antisymmetric requirement of Cooper pair wave functions,  the pairing matrices must satisfy $\hat{\Delta}^{T}(\bm{r})=-\hat{\Delta}(\bm{r})$.

 For simplicity, we only take momentum independent pairings into account.  There are only six matrices: $s_y$, $\tau_xs_y$, $\tau_zs_y$, $\tau_y$, $\tau_ys_x$, $\tau_ys_z$ can couple to the momentum independent pairings. The possible momentum-independent pairings in the layer and spin space can be classified in Table S1. The   $\Delta_{A_1,1}, \Delta_{A_2}$ are intralayer singlet pairings  and  the  $\Delta_{A_1,2},\Delta_{A1,3}$  are interlayer singlet pairings, while the $\Delta_{E_1},\Delta_{E_2}$ belonging to a two-dimensional irreducible representation is interlayer triplet pairings.   
 \begin{table}[h]
 	\caption{Classification of  possible  s-wave pairings for twisted bilayer TMD with the irreducible representations of $D_3$ point group in layer and spin space. Here $s, \tau$ are Pauli matrices defining in spin and layer space. }
 	\begin{tabular}{cccc}
 		\hline\hline
 		\multicolumn{2}{c}{Rep.} & Matrix form $\gamma$ &    Explicit form\\\hline
 		\multirow{2}{*}{$A_1$}&$\Delta_{A_1,1}$&$is_y$&$\psi_{-\xi
 			-\bm{k}, t\uparrow}\psi_{\xi+\bm{k}, t\downarrow}+\psi_{-\xi
 			-\bm{k},b\uparrow}\psi_{\xi+\bm{k},b\downarrow}$\\
 		&$\Delta_{A_1,2}$&$i\tau_xs_y$&$\psi_{-\xi-\bm{k},t\uparrow}\psi_{\xi+\bm{k},b\downarrow}+\psi_{-\xi-\bm{k},b\uparrow}\psi_{\xi+\bm{k},t\downarrow}$\\
 		&$\Delta_{A_1,3}$&$\tau_ys_x$&$\psi_{-\xi-\bm{k},t\uparrow}\psi_{\xi+\bm{k},b\downarrow}-\psi_{-\xi-\bm{k},b\uparrow}c_{\xi+\bm{k},t\downarrow}$\\
 		$A_2$&$\Delta_{A_2}$&$i\tau_zs_y$&$\psi_{-\xi-\bm{k},t\uparrow}\psi_{\xi+\bm{k},t\downarrow}-\psi_{-\xi-\bm{k}, b\uparrow}\psi_{\xi+\bm{k},b\downarrow}$\\
 		
 		\multirow{2}{*}{$E$}&	\multirow{2}{*}{\{$\Delta_{E_1},\Delta_{E_2}$\}}&	\multirow{2}{*}{\{$\tau_y$,$\tau_ys_z$\}}&\{$\psi_{-\xi-\bm{k},t\uparrow}\psi_{\xi+\bm{k},b\uparrow}+\psi_{-\xi-\bm{k},t\downarrow}\psi_{\xi+\bm{k},b\downarrow}$,\\
 		&&&$\psi_{-\xi-\bm{k},t\uparrow}\psi_{\xi+\bm{k},b\uparrow}-\psi_{-\xi-\bm{k}, t\downarrow}\psi_{\xi+\bm{k}, b\downarrow}$\}\\
 		\hline
 	\end{tabular}
 	\label{TableS1}
 \end{table}
 \newpage
 \section{Extended figures}
 
 \begin{figure}[h]
 	\centering
 	\includegraphics[width=0.7\linewidth]{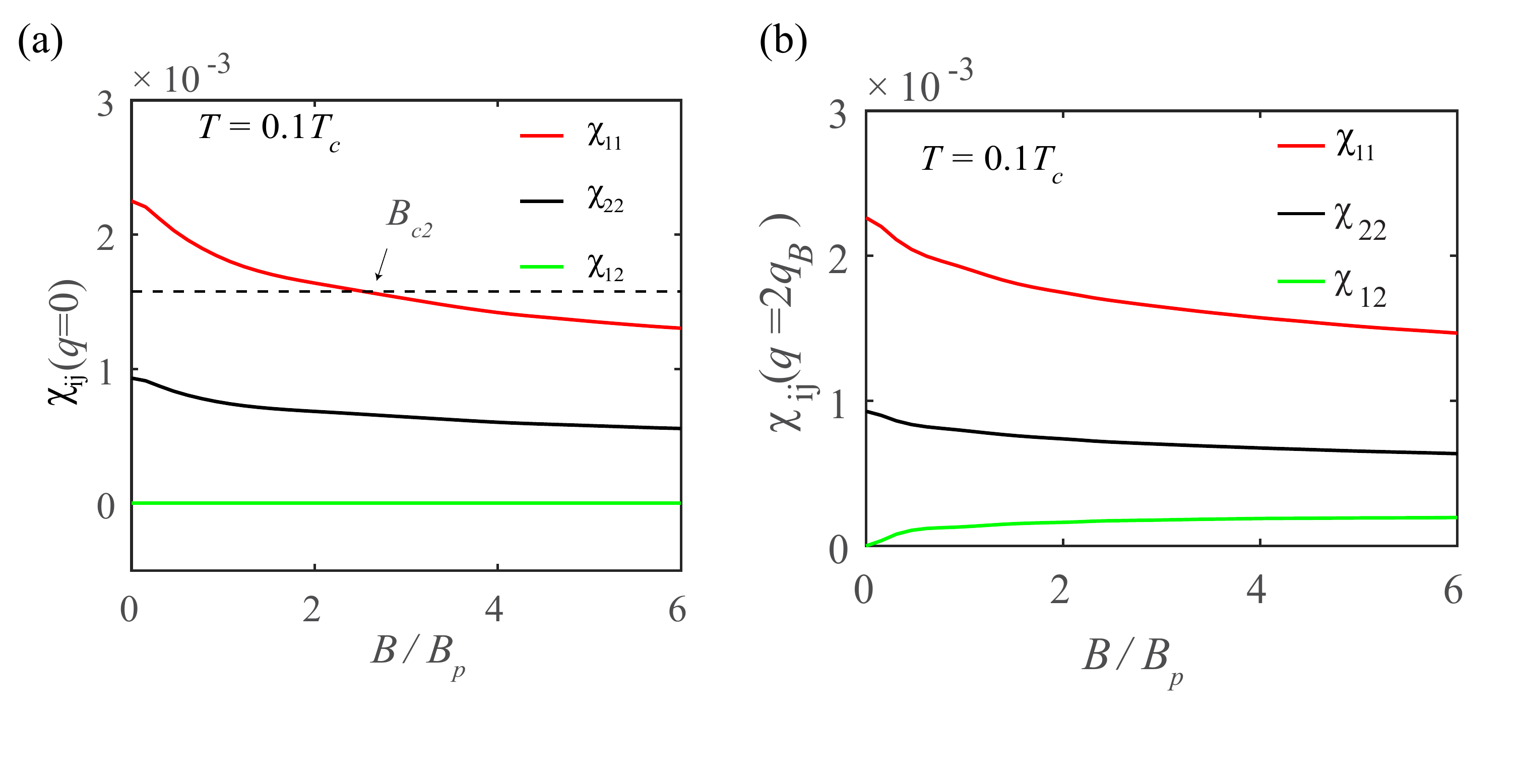}
 	\caption{Pairing susceptibility for $A_1$ pairing $\chi_{11}$, $A_2$ pairing, $\chi_{22}$ and their mixing $\chi_{12}$. (a) The zero-momentum pairing susceptibility $\chi_{ij}(q=0)$ versus in-plane magnetic fields $B$. The critical field $B_{c2}$ is highlighted. (b) The finite-momentum pairing susceptibility $\chi_{ij}(q=2q_B)$ versus $B$. Other parameters are the same as the main text Fig.~2.}
 	\label{fig:figs1}
 \end{figure}
 
 \begin{figure}[h]
 	\centering
 	\includegraphics[width=0.3\linewidth]{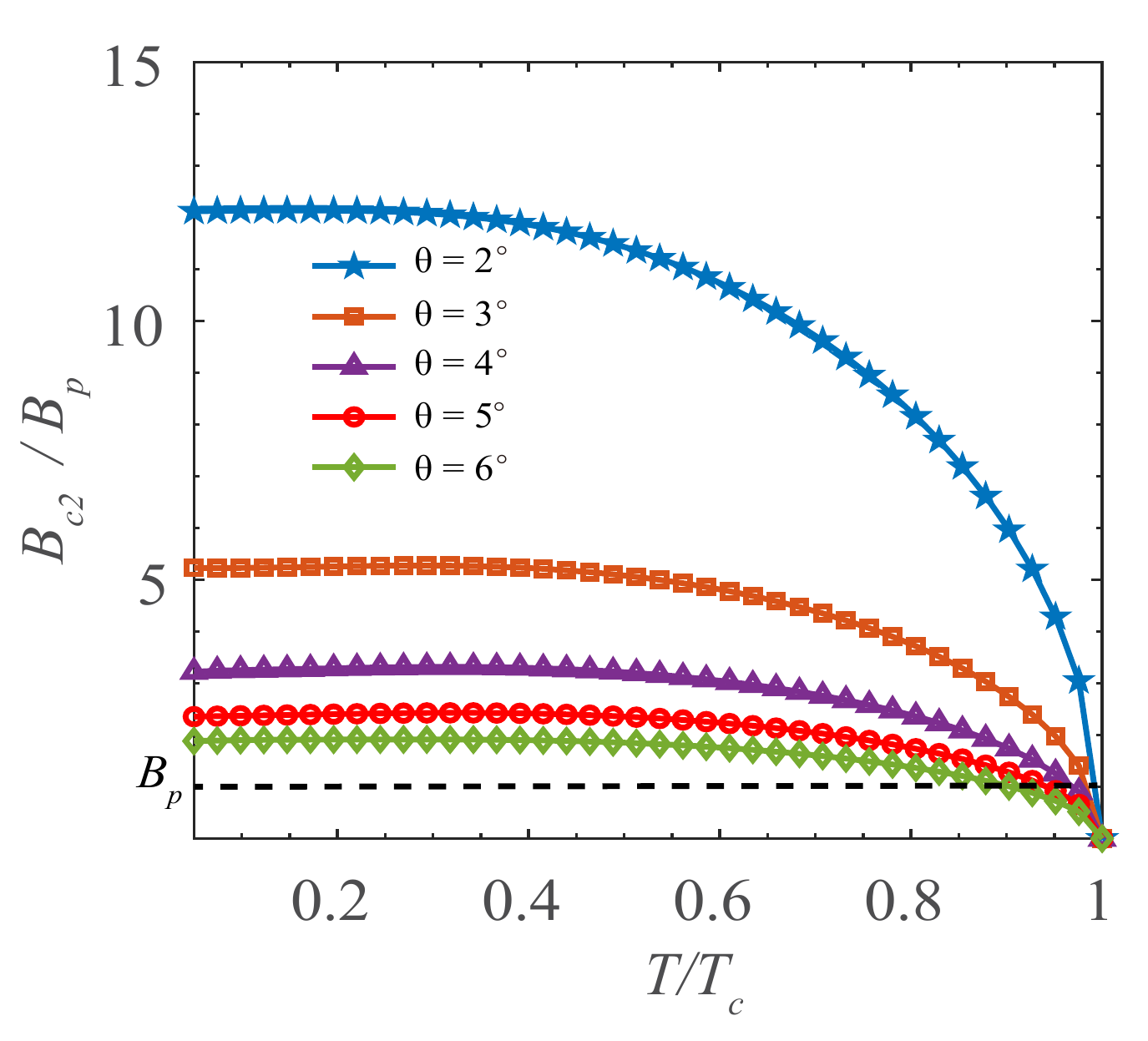}
 	\caption{The in-plane upper critical field $B_{c2}$ versus $T$ (with orbital effects) at various twist angle $\theta$.}
 	\label{fig:figs2}
 \end{figure}
 
 \begin{figure}[h]
 	\centering
 	\includegraphics[width=0.4\linewidth]{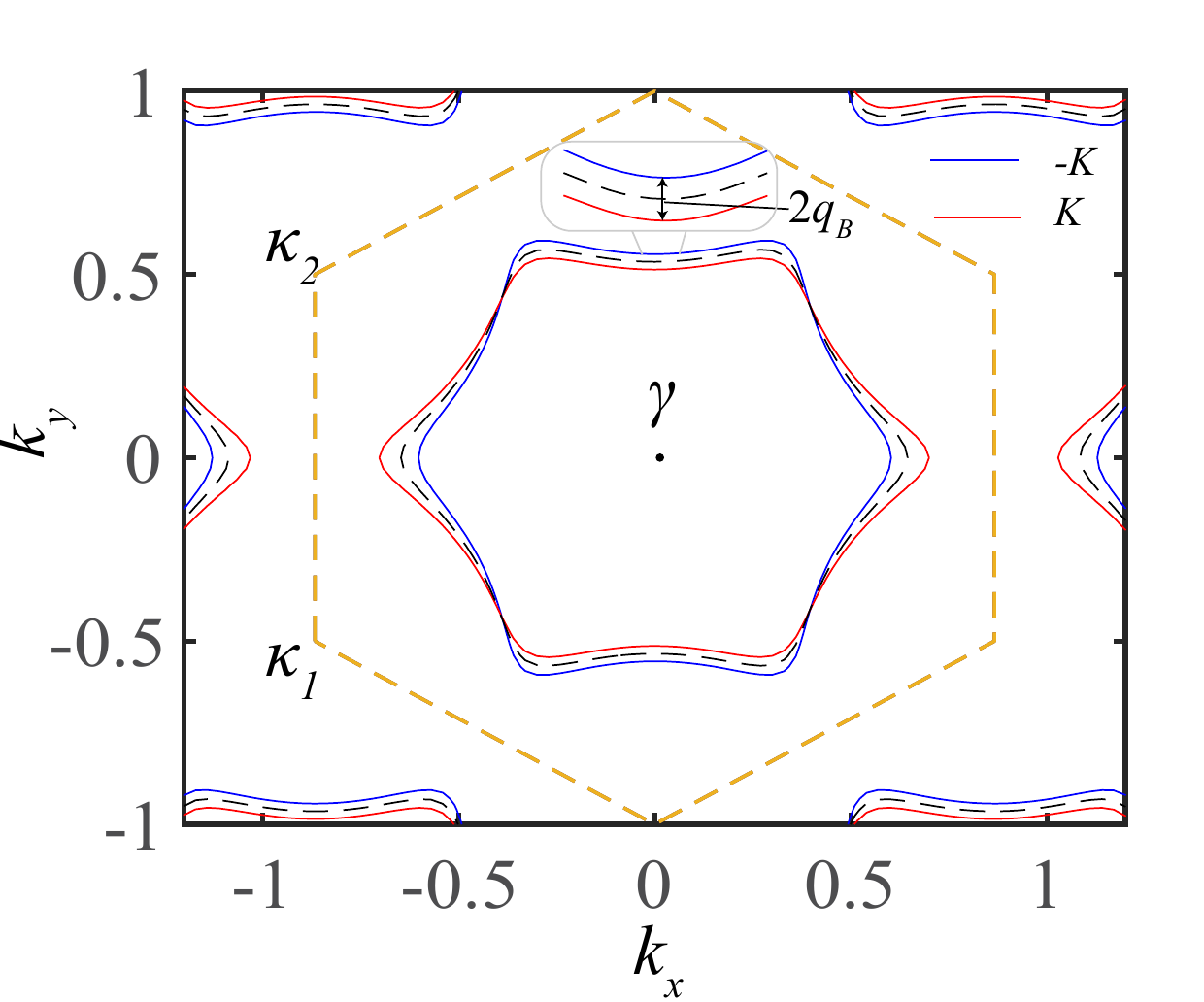}
 	\caption{The Fermi contour between at $+$ K valley (blue) and $-$ K valley (red) under finite magnetic field. The back dashed line labels the Fermi contour in the case without magnetic fields. The brown dashed line labels the boundaries of the Brillouin zone. To make the shifting of the Fermi contour more obvious,   we adopt a large in-plane magnetic field $B=20 B_p$.  The $2q_B$ momentum shifting between the Fermi contour of $K$ and $-K$ valley is highlighted (see the inset). Note that due to the interlayer hopping, the momentum shifting, in general, is not uniform.}
 	\label{fig:figs3}
 \end{figure}

 \newpage 
 \section{The linearized gap equation and free energy for the finite momentum pairing}
 
 As presented in the main text, the  mean-field Hamiltonian is written as  
 
 \begin{equation}
 	H_{MF}(\bm{r})=\sum_{\xi}\int d\bm{r}\Psi^{\dagger}_{\xi}(\bm{r})\mathcal{H}_{\xi}(\bm{r})\Psi_{\xi}(\bm{r})+\sum_{\xi}( \Psi^{\dagger}_{\xi }(\bm{r}) \hat{\Delta}(\bm{r}) \Psi^{\dagger}_{-\xi}(\bm{r})+\text{H.c.}).\label{model_m}
 \end{equation}
 Here,  the four-component annihilation  operator $\Psi_{\xi}(\bm{r})=(\psi_{\xi b\uparrow},\psi_{\xi b\downarrow},\psi_{\xi t\uparrow},\psi_{\xi t\downarrow})^{T}$. By directly transforming the continuum superconducting Hamiltonian  into momentum space, we obtain
 \begin{equation}
 	H_{MF}=\frac{1}{A}\sum_{\bm{p}}\Psi^{\dagger}_{\xi }(\bm{p})[H_0(\bm{p})]\Psi_{\xi }(\bm{p})+\frac{1}{A} \sum_{\bm{p'},\bm{q}}(\Psi^{\dagger}_{\xi}(\bm{p}+\frac{\bm{q}}{2}) \hat{\Delta} (\bm{q}) \Psi^{\dagger}_{-\xi}(-\bm{p}+\frac{\bm{q}}{2})+h.c.),
 \end{equation}
 where $A$ is the area of the moir\'e unit cell, $\bm{p}$ is the momentum within the first moir\'e Brillouin zone, $H_0(\bm{p})$ is the moir\'e Hamiltonian that can be  represented a plane wave basis, where the elements can be given by $\braket{\bm{p}+m\bm{g_2}+n\bm{g_3}|\mathcal{H}_{\xi}(\bm{r})|\bm{p}+m'\bm{g_2}+n'\bm{g_3}}$ with $m,n$ as integers, $\bm{g_j}$ as moir\'e wave vectors defined in the main text. The moir\'e  bands are obtained by diagonalizing the moir\'e Hamiltonian with a finite cut-off on $m,n$.

 \textit{The linearized gap equation.} We can decompose the pairings into the different channels  $\hat{\Delta}(\bm{q})=\sum_{i\mu}\Delta_{i\mu}(\bm{q})\gamma_{i\mu}$ with $\gamma_{i\mu}$ denoting the representation matrix defined in layer and spin space, and   
 the linearize gap equation is given by 
 \begin{equation}
 	\Delta_{i\mu}(\bm{q})=V_i\sum_{j\nu} \chi_{ij,\mu\nu}(\bm{q})\Delta_{j\nu}(\bm{q}),
 \end{equation}
 where $i,j$ label the representation, $\mu,\nu$ label the component in this representation, $V_i$ denotes the strength of attractive interaction,  $\bm{q}$ denotes the finite-momentum pairing of Cooper pairs.  The superconductivity susceptibility is given by
 \begin{equation}
 	\chi_{ij,\mu\nu}^{(2)}(\bm{q})=-\frac{1}{\beta}\sum_{i\omega_n,\bm{p}}\text{Tr}(\gamma_{i\mu}^{\dagger}G_{e}(\bm{p}+\bm{q}/2,i\omega_n)\gamma_{j\nu}G_h(\bm{p}-\bm{q}/2,i\omega_n)),
 	\label{eq:sus}
 \end{equation} 
 where $\beta=1/k_BT$,   the single-particle Green's functions for electrons $
 G_e(\bm{p},i\omega_n)=(i\omega_n-H_0(\bm{p}))^{-1}$ and holes $G_h(\bm{p},i\omega_n)=(i\omega_n+H^*_0(-\bm{p}))^{-1}$ with $H_0(\bm{p})$ as the moir\'e Hamiltonian.
 
 By utilizing the eigenstates of $H_0(\bm{p})$: $H_0(\bm{p})\ket{u_{a\bm{p}}}=E_{a}(\bm{p})\ket{u_{a\bm{p}}},
 H^{*}_0(\bm{p})\ket{\nu_{b\bm{p}}}=E'_{b}(\bm{p})\ket{\nu_{b\bm{p}}}$ ($a$ and $b$ are band indices), we can further simplify Eq.~\ref{eq:sus} as
 \begin{equation}
 	\chi_{ij,\mu\nu}^{(2)}(\bm{q})=\int_{\bm{p}} \sum_{a,b}O^{j\nu}_{a,b}(\bm{p},\bm{q})O^{i\mu*}_{a,b}(\bm{p},\bm{q})\mathcal{K}_{ab}(\bm{p},\bm{q},\bm{B}) 
 \end{equation}
 where the overlap function $O^{j\nu}_{a,b}(\bm{p},\bm{q})=\braket{u_{a\bm{p}+\bm{q}/2}|\gamma_{j\nu}|\nu_{b-\bm{p}+\bm{q}/2}}$ and  the kernel function
 \begin{equation}
 	\mathcal{K}_{ab}(\bm{p},\bm{q})=\frac{1-f(E_a(\bm{p}+\bm{q}/2))-f(E'_b(-\bm{p}+\bm{q}/2))}{E_{a}(\bm{p}+\bm{q}/2)+E'_{b}(-\bm{p}+\bm{q}/2)}
 \end{equation} 
 Here, $f$ is the Fermi distribution function. Note that in the calculation, it is sufficient to consider only the top moir\'e bands as the pairing energy scale is still much smaller than the moir\'e bandwidth.

 \textit{Free energy calculation.} In the main text, we have calculated the free energy of the FF pairing $\hat{\Delta}(\bm{r})=\Delta e^{i\bm{q}\cdot\bm{r}}i\sigma_y$.  Here, we present the detailed process. To be convenient, we perform a gauge  transform for the mean-field Hamiltonian Eq.~(\ref{model_m}): $\Psi^{\dagger}_{\xi}(\bm{r})\rightarrow \tilde{\Psi}^{\dagger}_{\xi}(\bm{r})e^{-i\frac{1}{2}\bm{q}\cdot\bm{r}}$. After this gauge transformation, we can obtain a Bogoliubov-de Genens (BdG) Hamiltonian to describe the FF pairing:
 \begin{equation}
 	H_{BdG}(\bm{p},\bm{q})=\begin{pmatrix}
 		H_0(\bm{p}+\frac{\bm{q}}{2})&\Delta i\sigma_y\\	(\Delta i\sigma_y)^{\dagger}&-H_0^{T}(-\bm{p}+\frac{\bm{q}}{2})
 	\end{pmatrix}
 \end{equation}
 The free energy at every finite momentum $\bm{q}$ can then be calculated with
 \begin{equation}
 	\mathcal{F}(\bm{q})=\frac{|\Delta|^2}{U_0}-\frac{1}{\beta}\sum_{\bm{p},n}\ln(1+e^{-\beta\epsilon_{\bm{p},n}(\bm{q})}).
 \end{equation}
 Here, $\epsilon_{\bm{p},n}(\bm{q})$ are the eigenenergies of $H_{BdG}(\bm{p},\bm{q})$,  the attractive interaction strength $U_0$  can be fixed by the critical temperature $T_c$.
 
 %As we will show, even when the interlayer interaction strength  $V_0$ is comparable to  $U_0$, the $\Delta_{A_1,2}$ interlayer pairing still can be neglected. 
 %
 %\section{The influence of the interlayer pairing $\Delta_{A_1,2}$}
 %
 %The in-plane magnetic field breaks $D_3$ symmetry such that  the $A_1$ pairing and $A_2$ pairing will mix with each other.  In principle,

 \section{Ginzburg-Landau Free energy for a bilayer superconductor under in-plane orbital magnetic fields}
 Phenomenologically, the Ginzburg-Landau (GL) free energy for a bilayer system under an in-plane orbital magnetic field can be written as
 \begin{align}
 	\mathcal{F}&=\mathcal{F}_c+\mathcal{F}_k+\mathcal{F}_J\label{free}\\
 	\mathcal{F}_c&=\frac{1}{2A}[\int d\bm{r}\sum_{l}(-\alpha_0)|\Delta_l(\bm{r})|^2+\frac{\beta}{2}|\Delta_{l}(\bm{r})|^4],\\ \mathcal{F}_k&=\frac{1}{2A}\int d\bm{r}\{\frac{1}{2m}\sum_l|\hat{\Pi}_l\Delta_{l}(\bm{r})|^2-\Gamma[(\Pi_t\Delta_{t})^*(\Pi_b\Delta_{b})+(\Pi_b\Delta_{b})^*(\Pi_t\Delta_{t})]\},\\
 	\mathcal{F}_J&=\frac{\lambda_J}{2A}\int d\bm{r}|\Delta_t(\bm{r})-\Delta_{b}(\bm{r})|^2
 \end{align}
 where $\Delta_{l}(\bm{r})$ is the order parameter layer $l$,  the canonical momentum $\hat{\Pi}_{l}=(-i\nabla+2e\bm{A}_l)$ with $\bm{A}_l=l\frac{d}{2}\bm{B}\times \hat{\bm{z}}$, $m=2m^*$ is the mass of Cooper pairings, $\lambda_J\propto \frac{e}{\hbar}N(0)t_c^2$ ($N(0)$ are the density of states near Fermi energy, $t_c$ represents the coupling strength)is the Josephson coupling energy between the two layers, $A$ is sample area.  Here,  $\mathcal{F}_c$ is the free energy saved by forming Cooper pairing, $\mathcal{F}_k$ contains kinetic energy arising from the  intralayer canonical momentum and interlayer canonical momentum mixing of  Cooper pairs,  $\Gamma$ denotes the  canonical momentum mixing strength between Cooper pair within two layers,  $\mathcal{F}_J$ describes the Josephson term which captures the interlayer pairing mixing. Note that due to the giant Ising SOC, we have neglected the paramagnetic free energy.

 Next, we simplify the free energy form in the following cases:
 
 (i) The case where the amplitude of the order parameter in each layer has no spatial dependence. In this case, the order parameter becomes $\Delta_{t}(\bm{r})\equiv|\Delta| e^{i\varphi_t }$ and  $\Delta_{b}(\bm{r})\equiv|\Delta| e^{i\varphi_b }$ , and the free energy is simplified as
 \begin{equation}
 	\mathcal{F} (|\Delta|)= -\alpha_0|\Delta|^2+\Lambda q_B^2|\Delta|^2+\frac{\beta_0}{2}|\Delta|^4+\lambda_J(1-\cos(\varphi_t-\varphi_{b}))|\Delta|^2,\label{uniform_momentum}
 \end{equation}
 which is presented in the main text as Eq.~(10). Here, $\Lambda=(4\Gamma+\frac{1}{m^*})$ The $A_1$ pairing with $\varphi_t=\varphi_b$ is thus more favorable so that 
 \begin{equation}
 	\mathcal{F} (|\Delta|)= -\alpha_0|\Delta|^2+\Lambda q_B^2|\Delta|^2+\frac{\beta_0}{2}|\Delta|^4.
 \end{equation} For the $A_1$ pairing, we estimate the critical magnetic field as 
 \begin{equation}
 	q_B^2=\frac{\alpha_0}{\Lambda}.\label{uniform}
 \end{equation}

 (ii) The case with layer-antisymmetric  FF pairing where $\Delta_{t}=|\Delta|e^{i\bm{q}\cdot\bm{r}}$ and $\Delta_{b}=|\Delta|e^{-i\bm{q}\cdot\bm{r}}$. Here, we have set $|\Delta_t|=|\Delta_b|$ to save $F_J$.  As discussed in the main text, the favored $\bm{q}=(0,2q_B)$. In this case, the free energy becomes
 \begin{equation}
 	\mathcal{F} (|\Delta|)= -\alpha_0|\Delta|^2+\frac{\beta_0}{2}|\Delta|^4+\lambda_J|\Delta|^2.
 \end{equation}
 It can be seen that the layered FF pairing would not pay kinetic energy but exhibit a finite Josephson energy $\lambda_J|\Delta|^2$. 
 
 (iii) The case with layer-symmetric FFLO pairing $\Delta_t=\Delta_b=|\Delta_+|e^{i\bm{q}\cdot \bm{r}}+|\Delta_-|e^{-i\bm{q}\cdot \bm{r}}$ with $\bm{q}=(0,2q_B)$, $|\Delta_{+}|^2+|\Delta_{-}|^2=|\Delta|^2$. Note that the pairing within the two layers is identical.
 In this case, the free energy is deduced as
 \begin{equation}
 	\mathcal{F} (|\Delta|)= -\alpha_0|\Delta|^2+\frac{\beta_0}{2}(|\Delta|^4+2|\Delta_+|^2|\Delta_{-}|^2)+\frac{2q_B^2}{m^*}|\Delta|^2.
 \end{equation}
 As $\beta_0>0$, the free energy is minimized with $|\Delta_+|=|\Delta|, |\Delta_{-}|=0$ or $|\Delta_-|=|\Delta|, |\Delta_{+}|=0$.  Hence, up to the fourth order of the free energy, the favored pairing can only take  $\Delta(\bm{r})=|\Delta|e^{i\bm{q}\cdot\bm{r}}$ or $\Delta(\bm{r})=|\Delta|e^{-i\bm{q}\cdot\bm{r}}$, which is the $2q_B$-FF pairing we study in the main text. Then, the free energy of this pairing is simplified as
 \begin{equation}
 	\mathcal{F} (|\Delta|)= -\alpha_0|\Delta|^2+\frac{\beta_0}{2}|\Delta|^4+\frac{2q_B^2}{m^*}|\Delta|^2.
 \end{equation}
 Notably, this 2$q_B$-FF pairing exhibits more intralayer kinetic energy but would not exhibit any kinetic energy from the Cooper canonical momentum mixing between two layers.   The critical magnetic field of the 2$q_B$-FF pairing is now given by 
 \begin{equation}
 	q_B^2=\frac{m^*\alpha_0}{2}.
 \end{equation}
 Therefore, this 2$q_B$-FF pairing  could survive at a higher magnetic field than the uniform pairings if 
 \begin{equation}
 	4m^*\Gamma>1.
 \end{equation}
 We can also infer the FF pairing would be more favorable than the LO pairing in the weak coupling where $\lambda_J\ll 2q_B^2/m^*$. 
 
 %To understand the phase transition between zero-momentum and finite momentum, we can take account of  the  $B$ field dependent of effective mass,  $m*\approx m_0+\lambda B^2$. As a result, 

 We  clarify here that the phenomenological free energy we present is to give a qualitative understanding of the results of the main text. Some relevant terms in the free energy can be different or some higher-order terms could play a role  in the realistic model of twisted bilayer TMDs. %Then we can capture the FFLO state with a purely phenomenological free energy
 %\begin{equation}
 %	-\alpha_0+\alpha_1q^2+\alpha_2(\bm{q}\times \bm{B})\cdot \hat{z}
 %\end{equation} 
 
 %It can be seen that comparing to the free energy of uniform pairings Eq.~(\ref{uniform_momentum}), the LO pairing does not exhibit Josephson energy.
 
 %Next, we consider the order parameter exhibits spatial dependence. In this case, we can write $\Delta_{l}(\bm{r})=\frac{1}{\sqrt{A}}\sum_{\bm{q_l}}\Delta_{l}(\bm{q_l})e^{i\bm{q_l}\cdot\bm{r}}$. We substitute $\Delta_{l}(\bm{r})$ into Eq.~(\ref{free}), and obtain
 %\begin{equation}
 %\mathcal{F}=\frac{1}{2A}\sum_{l,\bm{q_l}}(-\alpha_0+\frac{(\bm{q_l}+2l\bm{q_B})^2}{2m})|\Delta_{l}(\bm{q}_l)|^2+\frac{1}{2A}\lambda_J(\sum_{l,\bm{q}_l}|\Delta_{l}(\bm{q_l})|^2-\sum_{\bm{q}_{+},\bm{q}_{-}}(\Delta^*_{+}(\bm{q}_{+})\Delta_{-}(\bm{q_{-}})+h.c.)\delta_{\bm{q}_{+},\bm{q}_{-}})
 %\end{equation}
 
 \section{The layer anti-symmetric $2q_B$ FF pairing in  $AA$ stacking bilayer TMD in weak interlayer coupling limit}
 
 \subsection{Model}
 For the $AA$ stacking bilayer TMD without twisting under external magnetic fields $\bm{B}$, the effective low-energy Hamiltonian for valence bands is given by
 \begin{equation}
 	H_0(\bm{k}+\epsilon\bm{K})=-
 	\frac{(\hbar \bm{k}+ e\bm{A}\tau_z)^2}{2m^*}-\mu+\epsilon\beta_{so} s_z+t\tau_x. \label{Hamil17}
 \end{equation}
 where $\epsilon=\pm$ denote valley indices, $m^*$ is the effective mass of the valence bands of the monolayer TMD, $\mu$ is the chemical potential, $\beta_{so}$ is the Ising SOC strength, $t$ is the coupling strength between two TMD layers, and Pauli matrices $s_i$, $\tau_i$ operate on the spin-, layer-space, respectively. Notice this Hamiltonian breaks the inversion symmetry since the Ising SOC term $\epsilon\beta_{so} s_z$ is mapped to $-\epsilon\beta_{so} s_z$ under inversion operation.  The Zeeman effect from external fields is omitted by assuming a giant Ising SOC $\beta_{so} \gg u_B B$, while the orbital effect from external fields is included in the gauge potential $\bm{A}=\frac{d}{2}(\bm{B}\times \hat{\bm{z}})=\frac{dB}{2}(\sin\chi \hat{\bm{x}}-\cos\chi\hat{\bm{y}})$, which is  opposite for two layers. Here, $d$ denotes the interlayer separation, $\chi$ characterizes  the direction of the magnetic field. Inserting the chosen gauge potential into Hamiltonian (\ref{Hamil17}), we obtain
 \begin{eqnarray}
 	H_0(\bm{k}+\epsilon\bm{K})=-\frac{\hbar^2\bm{k}^2}{2m^*}-\frac{\hbar^2}{m^*}(k_B\sin\chi k_x-k_B\cos\chi k_y)\tau_z-\mu'+\epsilon\beta_{so} s_z+t\tau_x
 \end{eqnarray}
 with $l_0=\sqrt{\hbar/eB}, k_B=d/2l_0^2,\mu'=\mu+\hbar^2k_B^2/2m^*$. Then the eigenenergies are
 \begin{equation}
 	E_{\epsilon,s,\tau}(\bm{k})=-\frac{\hbar^2\bm{k}^2}{2m}-\mu'+\tau\sqrt{t^2+(\frac{\hbar^2k_B}{2m^*})^2(k_x^2\sin^2\chi+k_y^2\cos^2\chi-k_xk_y\sin2\chi)}+\epsilon s\beta_{so}.
 \end{equation}
 To be specific,   we set the magnetic field to be along the x direction ($\chi=0$)  in the following. The giant SOC can push some bands away from Fermi energy, and in this case, only the top valence bands  $E_{+,\uparrow,\tau}(\bm{k})$, $E_{-,\downarrow,\tau}(\bm{k})$ matter. By projecting the states on these top valence bands, we obtain an effective Hamiltonian as
 \begin{equation}
 	H_0(\bm{k}+\epsilon \bm{K})=-\frac{\hbar^2\bm{k}^2}{2m^*}-\mu-V(k_y)\tau_z+t\tau_x, \label{Hamil_eff}
 \end{equation}
 where the chemical potential $\mu$ is measured from the valence band top, the orbital field induced term $V(k_y)=\hbar v_B k_y$ with $v_B=\hbar k_B/m^*$. Notice the spin and valley are locked in this case: the $\bm{K}$ valley is locked as spin-up, while $-\bm{K}$ valley is locked as spin-down.  With the Hamiltonian (\ref{Hamil_eff}),  we can obtain 
 single-particle  Green's functions for normal states:
 \begin{eqnarray}
 	&G_e(\bm{k},i\omega_n)=(i\omega_n-H_0(\bm{k}))^{-1}=G_{+}(\bm{k},i\omega_n)+G_{-}(\bm{k},i\omega_n)\frac{-V\tau_z+t\tau_x}{\sqrt{t^2+V^2}}, \label{single_Green_1}\\ &G_h(\bm{k},i\omega_n)=(i\omega_n+H^*_0(-\bm{k}))^{-1}=-G_e^{T}(-\bm{k},-i\omega_n),
 	\label{single_Green_2}
 \end{eqnarray} 
 where $V\equiv V(k_y)$ for the compact of notation,  $G_{e (h)}(\bm{k},i\omega_n)$ is electron (hole) Green's function, and
 \begin{eqnarray}
 	G_{\pm}(\bm{k},i\omega)=\frac{1}{2}(\frac{1}{i\omega-\xi_{+}(\bm{k})}\pm\frac{1}{i\omega-\xi_{-}(\bm{k})})
 \end{eqnarray}
 with $\xi_{\pm}(\bm{k})=\xi_{\bm{k}}\pm  \sqrt{V^2+t^2}$, $\xi_{\bm{k}}=-\frac{\hbar^2\bm{k}^2}{2m}-\mu$, the Matsubara frequency $\omega_n=(2n+1)\pi k_BT$, and $T$ denotes the temperature.
 % $v_f=\hbar k_f/m$, $V=\frac{\hbar ev_fd B}{4\sqrt{2}}$.

 \subsection{Zero-momentum pairing}
 
 Within  in the Nambu basis $(\psi_{\bm{k}+\bm{K},t},\psi_{\bm{k}+\bm{K},b},\psi^{\dagger}_{-\bm{k}-\bm{K},t},\psi^{\dagger}_{-\bm{k}-\bm{K},b})$, it can be found the BDG Hamiltonian is written as
 \begin{equation}
 	H_{BDG}(\bm{k})=(\xi_{\bm{k}}+t\tau_x)\rho_z-V(k_y)\tau_z+\Delta\rho_x.
 \end{equation}
 Let us first consider the usual BCS zero-momentum pairing. For zero-momentum pairing, the superconductivity susceptibility is
 \begin{eqnarray}
 	\chi^{(2)}(\bm{q}=0)&&=-\frac{1}{\beta_{so}}\sum_{n,\bm{k}}\text{Tr}(G_{e}(\bm{k},i\omega_n)\tau_0G_h(\bm{k},i\omega_n)\tau_0)\nonumber\\
 	&&=\frac{2}{\beta_{so}}\sum_{n,\bm{k}}(G_{+}(\bm{k},i\omega_n)G_{+}(-\bm{k},-i\omega_n)+\frac{t^2-V^2}{t^2+V^2}G_{-}(\bm{k},i\omega_n)G_{-}(-\bm{k},-i\omega_n))\nonumber\\
 	&&=-\frac{1}{2\beta_{so}}\sum_{n,\bm{k}}(1+\frac{t^2-V^2}{t^2+V^2})(\frac{1}{i\omega_n-\xi_{\bm{k}}-D}\frac{1}{i\omega_n+\xi_{\bm{k}}+D}+\frac{1}{i\omega_n-\xi_{\bm{k}}+D}\frac{1}{i\omega_n+\xi_{\bm{k}}-D})+\nonumber\\
 	&&(1-\frac{t^2-V^2}{t^2+V^2})(\frac{1}{i\omega_n-\xi_{\bm{k}}+D}\frac{1}{i\omega_n+\xi_{\bm{k}}+D}+\frac{1}{i\omega_n-\xi_{\bm{k}}-D}\frac{1}{i\omega_n+\xi_{\bm{k}}-D})\label{eq_suscep}
 \end{eqnarray}
 Here $\beta_{so}=1/k_BT$, $D=\sqrt{V^2+t^2}$.
 It can be shown
 \begin{eqnarray}
 	\chi_0&&=-\frac{1}{\beta_{so}}\sum_{n,\bm{k}}\frac{1}{i\omega_n-\xi_{\bm{k}}+A}\frac{1}{i\omega_n+\xi_{\bm{k}}-B}\nonumber\\
 	&&=N_0\log(\frac{2e^{\gamma}\omega_D}{\pi k_BT})+N_0\psi(\frac{1}{2})-\frac{N_0}{2}(\psi(\frac{1}{2}-\frac{i(A-B)}{4k\pi k_BT})+\psi(\frac{1}{2}+\frac{i(A-B)}{4k\pi k_BT}))\label{eq_chi_0},
 \end{eqnarray}
 where $\omega_D$ is the Debye frequency, $N_0$ denote the density of states.
 %The following is the detail process, 
 %\begin{eqnarray}
 %\chi_0&&=-\frac{1}{\beta_{so}}\sum_{n,\bm{k}}\frac{1}{i\omega_n-\xi_{\bm{k}}+A}\frac{1}{i\omega_n+\xi_{\bm{k}}-B}\nonumber\\
 %&&=-\frac{1}{\beta_{so}}\sum_{n,\bm{k}}(\frac{1}{i\omega_n-\xi_{\bm{k}}+A}\frac{1}{i\omega_n+\xi_{\bm{k}}-B}-\frac{1}{i\omega_n-\xi_{\bm{k}}}\frac{1}{i\omega_n+\xi_{\bm{k}}})-\frac{1}{\beta_{so}}\sum_{n,\bm{k}}\frac{1}{i\omega_n-\xi_{\bm{k}}}\frac{1}{i\omega_n+\xi_{\bm{k}}}\nonumber\\
 %&&=N_0\log(\frac{2e^{\gamma}\omega_D}{\pi k_BT})-\frac{N_0}{\beta_{so}}\sum_n\int d\xi( \frac{1}{i\omega_n-\xi+A}\frac{1}{i\omega_n+\xi-B}-\frac{1}{i\omega_n-\xi}\frac{1}{i\omega_n+\xi})\\
 %&&=N_0\log(\frac{2e^{\gamma}\omega_D}{\pi k_BT})-\frac{N_0}{\beta_{so}}(\sum_{n\ge 0}\frac{-2\pi i}{2i\omega_n+A-B}-\frac{-2\pi i}{2i\omega_n}+\sum_{n<0}\frac{2\pi i}{2i\omega_n+A-B}-\frac{2\pi i}{2i\omega_n})\\
 %&&=N_0\log(\frac{2e^{\gamma}\omega_D}{\pi k_BT})+\frac{N_0}{\beta_{so}}\sum_{n\ge 0}(\frac{2\pi i}{2i\omega_n+A-B}-\frac{2\pi i}{2i\omega_n})+\frac{N_0}{\beta_{so}}\sum_{n\ge 0}(\frac{2\pi i}{2i\omega_n-(A-B)}-\frac{2\pi i}{2i\omega_n})\\
 %&&=N_0\log(\frac{2e^{\gamma}\omega_D}{\pi k_BT})+N_0\psi(\frac{1}{2})-\frac{N_0}{2}(\psi(\frac{1}{2}-\frac{i(A-B)}{4k\pi k_BT})+\psi(\frac{1}{2}+\frac{i(A-B)}{4k\pi k_BT})).
 %\end{eqnarray}
 With Eq.~\ref{eq_chi_0}, Eq.~\ref{eq_suscep} can be simplified as 
 \begin{equation}
 	\chi^{(2)}=2N_0\log(\frac{2e^{\gamma}\omega_D}{\pi k_BT})+ \frac{2\left\langle V^2\right\rangle }{t^2+\left\langle V^2\right\rangle }[N_0\psi(\frac{1}{2})-\frac{N_0}{2}(\psi(\frac{1}{2}-\frac{i\sqrt{\left\langle V^2\right\rangle +t^2}}{2\pi k_B T})+\psi(\frac{1}{2}+\frac{i\sqrt{\left\langle V^2\right\rangle +t^2}}{2\pi k_B T}))].
 \end{equation}
 Here $\left<V^2\right>=\frac{1}{2}\hbar^2v_B^2k_f^2$, $\left<...\right>$ denotes the averaging over Fermi surface.
 Therefore, the linearized gap equation is
 \begin{equation}
 	\log(\frac{T}{T_c})=\frac{\left<V^2\right>}{t^2+\left<V^2\right>}[\psi(\frac{1}{2})-\frac{1}{2}(\psi(\frac{1}{2}-\frac{i\sqrt{\left<V^2\right>+t^2}}{2\pi k_B T})+\psi(\frac{1}{2}+\frac{i\sqrt{\left<V^2\right>+t^2}}{2\pi k_B T}))].\label{linear_zero_momentum}
 \end{equation}
 
 When $\sqrt{\left<V^2\right>+t^2}\ll T_c$, near $T_c$, Eq.~\ref{linear_zero_momentum} gives
 \begin{equation}
 	B_c=\frac{8\pi k_B T_c}{ed v_f |\psi^{(2)}(1/2)|}\sqrt{1-\frac{T}{T_c}},
 \end{equation}
 where the Fermi velocity $v_f=\hbar k_f/m^*$.
 \subsection{2$q_B$ layer-antisymmetric FF pairing in weak coupling limit}
 
 Let us consider the finite-momentum pairing case. We show that the layer antisymmetric pairing momentum with $\bm{q}=(0, 2k_B)$ at one layer and $\bm{q}=(0, -2k_B)$ is at the other layer   is more favorable in the weak coupling limit.
 
 We assume the intra-layer pairing is dominant. As shown in Table~S1, there are intra-layer pairing channels: $A_1$ pairing and $A_2$ pairing. In the presence of an in-plane magnetic field, $\Delta_{A1,1}$ and $\Delta_{A2,1}$ will couple with each other. The Landau free energy, up to the second order, is given by
 \begin{equation}
 	F=\frac{1}{2}\sum_{\bm{q}}\begin{pmatrix}
 		\Delta_{A_1,1}^*&\Delta_{A2,1}^*
 	\end{pmatrix}\begin{pmatrix}
 		\frac{1}{U_0}-\chi_{11}(\bm{q})&-\chi_{12}(\bm{q})\\-\chi_{21}(\bm{q})&\frac{1}{U_0}-\chi_{22}(\bm{q})
 	\end{pmatrix}\begin{pmatrix}
 		\Delta_{A_1,1}\\
 		\Delta_{A_2,1}
 	\end{pmatrix},
 \end{equation}
 where $U_0$ denotes the intra-layer interaction strength and the superconductivity susceptibility $\chi$ can be written as
 \begin{equation}
 	\chi_{ij}(\bm{q})=-\frac{1}{\beta_{so}}\sum_{n,\bm{k}}\text{Tr}[\tau_i G_e(\bm{k}+\frac{\bm{q}}{2})\tau_j G_h(\bm{k}-\frac{\bm{q}}{2},i\omega_n)]
 \end{equation}
 with $\tau_1=\tau_0,\tau_2=\tau_z$. Substitute the single-particle Green's function Eq.~\ref{single_Green_1} and Eq.~\ref{single_Green_2}, the superconductivity susceptibility can further written as
 \begin{eqnarray}
 	\chi_{11}&=&\frac{2}{\beta_{so}} \sum_{n,\bm{k}}[G_{+}(\bm{k}+\frac{\bm{q}}{2},i\omega_n)G_{+}(-\bm{k}+\frac{\bm{q}}{2},-i\omega_n)+\frac{-V^2+t^2}{V^2+t^2}G_{-}(\bm{k}+\frac{\bm{q}}{2},i\omega_n)G_{-}(-\bm{k}+\frac{\bm{q}}{2},-i\omega_n)],\\
 	\chi_{12}&=&\frac{2}{\beta_{so}}\sum_{n,\bm{k}}\frac{V}{\sqrt{V^2+t^2}}[G_{+}(\bm{k}+\frac{\bm{q}}{2},i\omega_n)G_{-}(-\bm{k}+\frac{\bm{q}}{2},-i\omega_n)-G_{-}(\bm{k}+\frac{\bm{q}}{2},i\omega_n)G_{+}(-\bm{k}+\frac{\bm{q}}{2},-i\omega_n)],\\
 	\chi_{22}&=&\frac{2}{\beta_{so}}\sum_{n,\bm{k}}[G_{+}(\bm{k}+\frac{\bm{q}}{2},i\omega_n)G_{+}(-\bm{k}+\frac{\bm{q}}{2},-i\omega_n)-G_{-}(\bm{k}+\frac{\bm{q}}{2},i\omega_n)G_{-}(-\bm{k}+\frac{\bm{q}}{2},-i\omega_n)].
 \end{eqnarray}
 After some direct calculations, we obtain
 \begin{eqnarray}
 	\chi_{11}(\bm{q})&\approx&2N_0\log(\frac{2e^{\gamma}\omega_D}{\pi k_BT})+\frac{CN_0v_f^2q^2}{32\pi^2 k_B^2T^2}+\frac{CN_0\left<V^2\right>}{4\pi^2k_B^2T^2}\\
 	\chi_{12}(\bm{q})&\approx&-\frac{CN_0v_Bv_fk_fq_y}{8\pi^2k_B^2 T^2}\\
 	\chi_{22}(\bm{q})&\approx&2N_0\log(\frac{2e^{\gamma}\omega_D}{\pi k_B T})+\frac{CN_0v_f^2q^2}{32\pi^2 k_B^2T^2}+\frac{CN_0\left<V^2\right>+t^2}{4\pi^2k_B^2T^2},
 \end{eqnarray}
 where $\left<V^2\right>=\frac{1}{2}v_B^2k_f^2$, $C=\psi^{(2)}(\frac{1}{2})$. 
 
 In the weak coupling limit $t \rightarrow 0$, the Landau free energy is
 \begin{eqnarray}
 	F&\approx& \sum_{\bm{q}}\begin{pmatrix}
 		\Delta_{A_1,1}^*&\Delta_{A2,1}^*
 	\end{pmatrix}[(N_0\log(\frac{T}{T_c})-\frac{CN_0v_f^2q^2}{64\pi^2k_B^2T^2}-\frac{CN_0v_B^2k_f^2}{16\pi^2k_B^2T^2})\begin{pmatrix}
 		1&0\\0&1
 	\end{pmatrix}+\nonumber\\
 	&&\frac{CN_0v_Bv_fk_fq_y}{16\pi^2k_B^2 T^2}\begin{pmatrix}
 		0&1\\1&0
 	\end{pmatrix}]\begin{pmatrix}
 		\Delta_{A_1,1}\\
 		\Delta_{A_2,1}
 	\end{pmatrix}\nonumber\\
 	&=&\sum_{\bm{q}}\lambda_1|\Delta_1|^2+\lambda_2|\Delta_2|^2,
 \end{eqnarray}
 where $\Delta_{1,2}=\Delta_{A_1,1}\pm \Delta_{A_2,1}$, $\lambda_{1,2}=N_0\log(\frac{T}{T_{c}})-\frac{CN_0v_f^2q^2}{64\pi^2k_B^2T^2}-\frac{CN_0v_B^2k_f^2}{16\pi^2k_B^2T^2}\pm\frac{CN_0v_Bv_fk_fq_y}{16\pi^2k_B^2 T^2}$. The critical temperature is given by $\min(\lambda_1,\lambda_2)=0$. This gives 
 \begin{equation}
 	N_0\log(\frac{T}{T_{c}})-\frac{CN_0}{16\pi^2 k_B^2 T_c^2}[(\frac{q_yv_f}{2}-\text{sgn}(q_y)v_Bk_f)^2+\frac{q_x^2v_f^2}{4}]=0.
 \end{equation}
 The critical temperature $T_c$ is maximized when $q_x=0, q_y=\text{sgn}(q_y)2v_B k_f/v_f$. It is easy to see $q_y=\pm2k_B$ with $v_B=\hbar k_B/m, v_f=\hbar k_f/m$. And hence, in weak coupling limit $t\rightarrow 0$, the finite-momentum pairing  $\Delta_1(\bm{q})=\left< c^{\dagger}_{\bm{K}+\bm{k}+\bm{q}/2,+}c^{\dagger}_{-\bm{K}-\bm{k}+\bm{q}/2,+}\right>$ with $\bm{q}=(0, 2k_B)$ and $\Delta_2(\bm{q})=\left< c^{\dagger}_{\bm{K}+\bm{k}+\bm{q}/2,-}c^{\dagger}_{-\bm{K}-\bm{k}+\bm{q}/2,-}\right>$ with $\bm{q}=(0, -2k_B)$ are stabilized. 
	
\end{document}